\documentclass[usenatbib]{mnras}
\usepackage{amsfonts,amsmath,amssymb,bm,ctable,verbatim,ulem,newtxtext,newtxmath}

\newcommand{\etal}{et al.}
\newcommand{\msun}{M_{\sun}}
\newcommand{\altaffilmark}[1]{$^{#1}$}
\newcommand{\altaffiltext}[1]{\altaffilmark{#1}}
\newcommand{\acknowledgments}[1]{\begin{small}\section*{Acknowledgments}\end{small}{\noindent #1}\vspace{5pt}}
\newcommand{\dataavailability}[1]{\begin{small}\section*{Data Availability}\end{small}{\noindent #1}\vspace{5pt}}
\newcommand{\ICsurl}{\href{http://www.tapir.caltech.edu/~phopkins/publicICs}{\url{http://www.tapir.caltech.edu/~phopkins/publicICs}}}
\newcommand{\FIREurl}{\href{http://fire.northwestern.edu}{\url{http://fire.northwestern.edu}}}
\newcommand{\gizmourl}{\href{http://www.tapir.caltech.edu/~phopkins/Site/GIZMO.html}{\url{http://www.tapir.caltech.edu/~phopkins/Site/GIZMO.html}}}
\newcommand{\HI}{H{\sc ~i}\ }

\newcommand{\NV}{N{\sc ~v}\ }

\newcommand{\OVI}{O{\sc ~vi}\ }
\newcommand{\OVII}{O{\sc ~vii}\ }

\newcommand{\MgII}{Mg{\sc ~ii}\ }
\newcommand{\SiIV}{Si{\sc ~iv}\ }
\newcommand{\NeVIII}{Ne{\sc ~viii}\ }

\definecolor{darkgreen}{RGB}{0,100,0}

\newcommand{\delmath}[1]{{\color{red}{\ifmmode\text{\sout{\ensuremath{#1}}}\else\sout{#1}\fi}}}

\title[CGM in cosmic ray-dominated galaxy halos]{Properties of the Circumgalactic Medium in Cosmic Ray-Dominated Galaxy Halos}
\author[Ji \etal]{
\parbox[t]{\textwidth}{
Suoqing Ji,\altaffilmark{1}
T.~K.\ Chan,\altaffilmark{2}
Cameron B.~Hummels,\altaffilmark{1}
Philip F.~Hopkins,\altaffilmark{1}
Jonathan Stern,\altaffilmark{4}
Du\v{s}an Kere\v{s},\altaffilmark{2}
Eliot Quataert,\altaffilmark{3}
Claude-Andr{\'e} Faucher-Gigu{\`e}re\altaffilmark{4} and
Norman Murray\altaffilmark{5}
} \vspace*{4pt} \\
\altaffiltext{1}{TAPIR, Mailcode 350-17, California Institute of Technology, Pasadena, CA 91125, USA. E-mail:suoqing@caltech.edu} \\
\altaffiltext{2}{Department of Physics, Center for Astrophysics and Space Science, University of California at San Diego, La Jolla, CA 92093} \\ 
\altaffiltext{3}{Department of Astronomy and Theoretical Astrophysics Center, University of California Berkeley, Berkeley, CA 94720} \\ 
\altaffiltext{4}{Department of Physics and Astronomy and CIERA, Northwestern University, 2145 Sheridan Road, Evanston, IL 60208, USA} \\
\altaffiltext{5}{Canadian Institute for Theoretical Astrophysics, 60 St George Street, University of Toronto, ON M5S 3H8, Canada}
}

\date{}
\begin{document}
\maketitle
\label{firstpage}

\begin{abstract}
We investigate the impact of cosmic rays (CRs) on the circumgalactic medium (CGM) in FIRE-2 simulations, for ultra-faint dwarf through Milky Way (MW)-mass halos hosting star-forming (SF) galaxies. Our CR treatment includes injection by supernovae, anisotropic streaming and diffusion along magnetic field lines, collisional and streaming losses, with constant parallel diffusivity $\kappa\sim3\times10^{29}\,\mathrm{cm^2\ s^{-1}}$ chosen to match $\gamma$-ray observations. With this, CRs become more important at larger halo masses and lower redshifts, and dominate the pressure in the CGM in MW-mass halos at $z\lesssim 1-2$. The gas in these ``CR-dominated'' halos differs significantly from runs without CRs: the gas is primarily cool (a few $\sim10^{4}\,$K), and the cool phase is volume-filling and has a thermal pressure below that needed for virial or local thermal pressure balance. Ionization of the ``low'' and ``mid'' ions in this diffuse cool gas is dominated by photo-ionization, with \OVI columns $\gtrsim 10^{14.5}\,\mathrm{cm^{-2}}$ at distances $\gtrsim 150\,\mathrm{kpc}$. CR and thermal gas pressure are locally anti-correlated, maintaining total pressure balance, and the CGM gas density profile is determined by the balance of CR pressure gradients and gravity. Neglecting CRs, the same halos are primarily warm/hot ($T\gtrsim 10^{5}\,$K) with thermal pressure balancing gravity, collisional ionization dominates, \OVI columns are lower and \NeVIII higher, and the cool phase is confined to dense filaments in local thermal pressure equilibrium with the hot phase.
\end{abstract}


\section{Introduction}
\label{sect:intro}

Galaxies are not isolated systems; instead, they are embedded in extended dark matter and gaseous halos. Early models assumed that massive halos were filled with uniform, ``hot'' ($T\gtrsim10^6\ \mathrm{K}$) virialized gas, predicted by hydrostatic equilibrium arguments  \citep{bahcall1969absorption,white1978core}, and the existence of such gas in massive gaseous halos has been supported by observations of X-ray emission \citep{li2008chandra,fang2012hot}, \OVII line absorption \citep{wang2005warm} and the Sunyaev-Zel'dovich (SZ) effect \citep{planck13,anderson15}. While at high redshift cold filamentary gas is expected to survive even in relatively massive halos (e.g., \citealt{kerevs2005galaxies,dekel2006galaxy}), at low redshift, colder filaments are expected to be heated up in halos hosting $L_{\ast}$ galaxies. However, the combined mass of the hot halo gas and the galaxy disks in MW-mass and smaller halos falls below the expected Universal baryon fraction predicted in simple structure formation models \citep{mcgaugh2009baryon,miller2013structure}. This discrepancy is partially relieved by the recent discoveries of cooler component in the halo gas detected via quasar absorption lines at low redshifts \citep{stocke13,werk2014cos,prochaska2017cos}, and via Ly$\alpha$ emission at high redshifts \citep{cantalupo14, hennawi15, cai17}. The co-existence of both cool and hot phases has led to a new picture in which galaxies are surrounded by multiphase CGM gas extending from the disk ($\sim 10\,$kpc) to the virial radius ($\sim 300\,$kpc), which serves as a reservoir containing most of the baryons and potentially playing a critical role in ``feedback'' processes critical to galaxy formation \citep{tumlinson2017circumgalactic,zhang2018review}.

These and other observations raise many unsolved questions. For example, cool photo-ionized gas at a few $10^4\ \mathrm{K}$ traced by ``low'' ions\footnote{Throughout, we follow convention and refer to {ions with ionization energy $E_\mathrm{i} \lesssim 40\,\mathrm{eV}$ ($T=10^{4-4.5}\,\mathrm{K}$)} as ``low'' ions, {$40\,\mathrm{eV}\lesssim E_\mathrm{i}\lesssim 100\,\mathrm{eV}$ ($T=10^{4.5-5.5}\,\mathrm{K})$} as ``mid'' ions, and {$E_\mathrm{i} \gtrsim 100\,\mathrm{eV}$ ($T>10^{5.5}\,\mathrm{K}$)} as ``high'' ions \citep{tumlinson2017circumgalactic}.} has been found to have an electron density at least an order-of-magnitude lower than that expected if it were in thermal pressure equilibrium with the hot phase -- it appears to be ``under-pressured'' \citep{werk2014cos,mcquinn2018implications}. How the cool phase embedded in a hot medium could form and survive disruption via fluid-mixing instabilities remains unclear. In \OVI (a ``mid'' ion), large columns $\sim10^{14.5}\ \mathrm{cm^{-2}}$ are observed to distances $\gtrsim 150\ \mathrm{kpc}$ from star-forming galaxies  \citep{tumlinson2011large,werk2016cos}. The nature and origin of the \OVI is still debated: it could be collisionally-ionized warm gas at $10^{5.5}\ \mathrm{K}$, originating in turbulent mixing \citep{begelman1990turbulent,kwak2010numerical,voit2018ambient} or thermal conduction \citep{gnat2010metal} layers (as it is thermally unstable), or it could trace the cool low-density photo-ionized gas \citep{stern2018does}. The observed \OVI characteristic distance of $\gtrsim 150 \, \mathrm{kpc}$ has also posed a challenge for recent theoretical CGM models (e.g., \citealt{mathews2017circumgalactic,faerman2017massive,stern2018does,stern2019cooling}). In numerical simulations, the column densities of the low and mid ions in CGM are usually under-predicted (e.g., \citealt{hummels2013constraints,cen2013composition,oppenheimer2016bimodality,liang2016column,fielding2016impact}), even in recent simulations which ``zoom in'' on the CGM to dramatically improve numerical resolution  \citep{peeples2018figuring,hummels2018impact,van2018cosmological}. 

These discrepancies could indicate an essential piece is missing from many models of the CGM, in the form of non-thermal components such as magnetic fields and/or cosmic rays (CRs). In the CGM, magnetic fields can facilitate cool gas formation by enhancing thermal instability \citep{ji2018impact}, protect the cool gas against hydrodynamic instabilities via magnetic tension \citep{dursi2008draping,mccourt2015magnetized}, and regulate anisotropic thermal conduction \citep{su2017feedback,su2019failure}, but most studies have concluded that these effects are relatively weak and do not qualitatively change the phase balance of the CGM \citep[see e.g.][]{komarov2014suppression,su2017feedback,hopkins2019but,su2019failure}. On the other hand, CR pressure could be responsible for supporting the diffuse cool CGM \citep{salem2016role,butsky2018role}, driving galactic outflows \citep{ruszkowski2017global,wiener2017interaction,chan2019cosmic}, or heating the CGM via excitation of short-wavelength Alfv\'en waves \citep{wiener2013cosmic}, and its non-linear effects on CGM structure in a fully cosmological setting remain largely unexplored. Especially, \citet{salem2016role} found that CRs can provide significant pressure support for CGM, while they did not include non-adiabatic CR energy loss terms which are very important in regimes of low CR diffusion coefficients or high gas densities (e.g., in central galaxies).

To explore the impact of the non-thermal components on the CGM in a more self-consistent manner, we utilize a new series of high-resolution, fully-cosmological simulations from the Feedback in Realistic Environments (FIRE)\footnote{FIRE project website: \FIREurl} project \citep{hopkins2014galaxies},  including magnetic fields, physical conduction and viscosity, and explicit CR transport and CR-gas interactions including collisional (hadronic+Coulomb) and streaming losses of CR energy \citep{chan2019cosmic,su2019failure,hopkins2019but}. Previous FIRE simulations, ignoring explicit CR transport, have been used to explore and predict CGM properties such as high-redshift HI covering factors \citep{faucher2015neutral,faucher2016stellar}, the nature of the cosmic baryon cycle and outflow recycling \citep{muratov2015gusty,angles2017cosmic,hafen2018origins}, statistics of low-redshift Lyman limit systems \citep{hafen2017low}, galaxy outflow properties and the metal budget of the CGM \citep{ma2015origin,muratov2017metal}, the SZ effect and halo baryon fractions \citep{van2016impact}, and temperature/density/entropy profiles of massive halos and clusters \citep{su2019failure}. We therefore extend these by considering the role of CRs. 

In \S\ref{sect:methods}, we briefly review the simulations and numerical methods. \S\ref{sect:results} presents and analyzes CGM properties, focusing on the effects of CRs where they significantly influence our predictions, including the CGM phase structure and direct comparisons with observed low/medium/high ion column density measurements. In \S\ref{sect:itp}, we provide a simple theoretical models for our simulation results, and in \S\ref{sect:diss} we summarize and discuss caveats of this work.

\section{Methods}
\label{sect:methods}

\begin{footnotesize}
\ctable[
  caption={{\normalsize Zoom-in simulations studied here (see \citealt{hopkins2019but} for details). All units are physical.}\label{tbl:sims}},center,star]{lccccccr}{
\tnote[ ]{Properties listed refer only to the ``target'' halo around which the high-resolution volume is centered, 
{\bf (1)} Simulation Name: Designation used throughout this paper. 
{\bf (2)} Final Redshift: Redshift to which the simulations are run.
{\bf (3)} $M_{\rm halo}^{\rm vir}$: Virial mass \citep[following][]{bryan.norman:1998.mvir.definition} of the ``target'' halo.
{\bf (4)} $M_{\ast}^{\rm MHD+}$: Central galaxy stellar mass, in our non-CR, but otherwise full-physics ``MHD+'' run. 
{\bf (5)} $M_{\ast}^{\rm CR+}$: Stellar mass, in our default CR+ ($\kappa_{\|}=3e29$) run.
{\bf (6)} $m_{i,\,1000}$: Baryonic (gas or star) particle mass, in units of $1000\,\msun$. The DM particle mass is larger by the universal ratio. 
{\bf (7)} $\langle \epsilon_{\rm gas} \rangle^{\rm sf}$: Gravitational force softening (Plummer-equivalent) at the mean density of star formation (gas softenings are adaptive to match hydrodynamic resolution). 
{\bf (8)} Additional notes.
All properties are measured at $z=0$, except for halo {\bf m12z2}, where all properties are measured at $z=2$ (after it reaches MW masses).
\vspace{-0.5cm}
}
}{
\hline\hline
Simulation & $z_{\rm final}$ &  $M_{\rm halo}^{\rm vir}$ & $M_{\ast}^{\rm MHD+}$ & $M_{\ast}^{\rm CR+}$ & $m_{i,\,1000}$ & $\langle \epsilon_{\rm gas} \rangle^{\rm sf}$ & Notes \\
Name \, &  & $[\msun]$ &  $[\msun]$  &   $[\msun]$  &$[1000\,\msun]$ & $[{\rm pc}]$ & \, \\ 
\hline 
{\bf m10q} & 0 & 8.0e9 & 2e6 & 2e6 & 0.25 & 0.8 & isolated dwarf in an early-forming halo \\
{\bf m11b} & 0 &  4.3e10 & 8e7 & 8e7 & 2.1 & 1.6 & disky (rapidly-rotating) dwarf \\
{\bf m11f} & 0 &  5.2e11 & 3e10 & 1e10 & 12 & 2.6 & early-forming, intermediate-mass halo \\
{\bf m12i} & 0 &  1.2e12 & 7e10 & 3e10 & 7.0 & 2.0 &  late-forming, MW-mass with massive disk\\  
{\bf m12m} & 0 &  1.5e12 & 1e11 & 3e10 & 7.0 & 2.3 & earlier-forming halo, features strong bar \\ 
{\bf m12z2} & 2 &  1.5e12 & 2e11 & 2e11 & 56 & 2.3 & massive halo at high-$z$: properties at $z=2$ \\ 
\hline\hline
}
\end{footnotesize}

The specific simulations studied here are the same as those presented and studied in \citet{hopkins2019but}, where the details of the numerical methods are described. We therefore only briefly summarize here. The simulations were run with {\small GIZMO}\footnote{A public version of {\small GIZMO} is available at \gizmourl} \citep{hopkins:gizmo}, in its meshless finite-mass MFM mode (a mesh-free finite-volume Lagrangian Godunov method). The simulations solve the equations of ideal magneto-hydrodynamics (MHD) as described and tested in \citep{hopkins:mhd.gizmo,hopkins:cg.mhd.gizmo}, with anisotropic Spitzer-Braginskii conduction and viscosity as described in \citet{hopkins:gizmo.diffusion,su2017feedback} and \citet{hopkins2019but}. Gravity is solved with adaptive Lagrangian force softening (so hydrodynamic and force resolutions are matched). The simulations are fully-cosmological ``zoom-in'' runs with a high-resolution region (of size ranging from $\sim 1$ to a few Mpc on a side) surrounding a ``primary'' halo of interest \citep{onorbe:2013.zoom.methods};\footnote{For the MUSIC \citep{hahn:2011.music.code.paper} files necessary to generate all ICs here, see:\\ \ICsurl} the properties of these primary halos (our main focus here, as these are the best-resolved in each box) are given in Table~\ref{tbl:sims}.

All our simulations include the physics of cooling, star formation, and stellar feedback from the FIRE-2 code, described in detail in \citet{hopkins:fire2.methods}. Gas cooling is followed from $T=10-10^{10}\,$K (including a variety of process, e.g.\ metal-line, molecular, fine-structure, photo-electric, photo-ionization, and more, accounting for self-shielding and both local radiation sources and the meta-galactic background; see \citealt{hopkins:fire2.methods}), and the \citet{faucher2009new} UV background (hereafter FG09) is adopted. We follow 11 distinct abundances accounting for turbulent diffusion of metals and passive scalars as in \citet{colbrook:passive.scalar.scalings,escala:turbulent.metal.diffusion.fire}. Gas is converted to stars using a sink-particle prescription if and only if it is locally self-gravitating at the resolution scale \citep{hopkins:virial.sf}, self-shielded/molecular \citep{krumholz:2011.molecular.prescription}, Jeans-unstable, and denser than $>1000\,{\rm cm^{-3}}$. Each star particle is then evolved as a single stellar population with IMF-averaged feedback properties calculated following \citep{starburst99} for a \citet{kroupa:2001.imf.var} IMF and its age and abundances. We explicitly treat mechanical feedback from SNe (Ia \&\ II) and stellar mass loss (from O/B and AGB stars) as discussed in \citet{hopkins:sne.methods}, and radiative feedback including photo-electric and photo-ionization heating and UV/optical/IR radiation pressure with a five-band radiation-hydrodynamics scheme as discussed in \citet{hopkins:radiation.methods}. 

Magnetic fields, anisotropic Spitzer-Braginskii conduction and viscosity are included in our simulations as well. Conduction adds the parallel heat flux $\kappa_{\rm cond}\,\hat{\bm{B}}\,(\hat{\bm{B}}\cdot \nabla T)$, while viscosity adds the anisotropic stress tensor $\Pi \equiv -3\,\eta_{\rm visc}\,(\hat{\bm{B}}\otimes\hat{\bm{B}} - \mathbb{I}/3)\,(\hat{\bm{B}}\otimes\hat{\bm{B}} - \mathbb{I}/3) : (\nabla\otimes{\bm{v}})$ to the gas momentum and energy equations. The parallel transport coefficients $\kappa_{\rm cond}$ and $\eta_{\rm visc}$ follow the usual \citet{spitzer:conductivity,braginskii:viscosity} form, accounting for saturation following \citet{cowie:1977.evaporation}, and accounting for plasma instabilities (e.g.\ Whistler, mirror, and firehose) limiting the heat flux and anisotropic stress at high plasma-$\beta$ following \citet{komarov:whistler.instability.limiting.transport,squire:2017.max.braginskii.scalings,squire:2017.kinetic.mhd.alfven,squire:2017.max.anisotropy.kinetic.mhd}. 

Our ``CRs'' or ``CR+'' simulations include all of the above, and add our ``full physics'' treatment of CRs as described in detail in \citet{chan2019cosmic} and \citet{hopkins2019but}. We evolve a ``single bin'' ($\sim$\,GeV) or constant spectral distribution of CRs as an ultra-relativistic ($\gamma=4/3$) fluid, accounting for injection in SNe shocks (with a fixed fraction $\epsilon_{\rm cr}=0.1$ of the initial SNe ejecta kinetic energy in each time-resolved explosion injected into CRs), collisional (hadronic and Coulomb) losses from the CRs (with a fraction of this loss thermalizing and heating gas) following \citet{guo.oh:cosmic.rays}, advection and adiabatic work (in the local ``strong coupling'' approximation, so the CR pressure contributes to the total pressure in the Riemann problem for the gas equations-of-motion), and CR transport including anisotropic diffusion and streaming \citep{mckenzie.voelk:1982.cr.equations}. We solve the transport equations using a two-moment approximation to the full collisionless Boltzmann equation (similar to the schemes in \citealt{jiang.oh:2018.cr.transport.m1.scheme,thomas2019cosmic}), with a constant parallel diffusivity $\kappa_{\|}$ (perpendicular $\kappa_{\bot}=0$); streaming velocity ${\bm{v}}_{\rm stream} = -v_{\rm stream}\,\hat{\bm{B}}\,(\hat{\bm{B}}\cdot \hat{\nabla} P_{\rm cr})$ with $v_{\rm stream}=v_{A}$, the local Alfv\'en speed \citep{skilling:1971.cr.diffusion,holman:1979.cr.streaming.speed,kulsrud:plasma.astro.book,yan.lazarian.2008:cr.propagation.with.streaming}; and the ``streaming loss'' term ${\bm{v}}_{\rm stream}\cdot \nabla P_{\rm cr}$ thermalized (representing losses to plasma instabilities at the gyro scale; \citealt{wentzel:1968.mhd.wave.cr.coupling,kulsrud.1969:streaming.instability}). 

Our ``baseline'' or ``no CRs'' simulations include all the physics above except CRs: these are the ``MHD+'' simulations in \citet{hopkins2019but}. Note there we also compared a set without magnetic fields, conduction, or viscosity (the ``Hydro+'' runs); but as shown therein and in \citet{su2017feedback}  the differences in these runs are largely negligible, and we confirm this here. 
Our default ``CR'' simulations adopt $\epsilon_{\rm cr}=0.1$, $v_{\rm stream}=v_{A}$, and $\kappa_{\|}=3\times10^{29}\,{\rm cm^{2}\,s^{-1}}$, along with the full CR transport physics including anisotropic streaming, diffusion, collisional losses, etc.: these are the ``CR+($\kappa=3e29$)'' simulations in \citet{hopkins2019but}. Although we considered variations to all of these CR physics and, in particular, the diffusivity (which is not known to an order of magnitude, e.g., \citealt{zweibel2013microphysics,grenier2015nine}), in \citet{hopkins2019but,chan2019cosmic}, we showed that the observational constraints from e.g.\ spallation and more detailed measurements in the MW and $\gamma$-ray emission in local galaxies were all consistent with the default ($\kappa_{\|}=3\times10^{29}\,{\rm cm^{2}\,s^{-1}}$) model here, and ruled out models (within the context of the approximations here) with much lower/higher $\kappa_{\|}$ (or $v_{\rm stream}$).\footnote{See \citet{chan2019cosmic} for detailed discussions regarding the constrain of the CR diffusion coefficient. It is also worth noting that some studies (e.g., \citealt{girichidis2018cooler}) quote a lower diffusion coefficient ($3\times10^{28}\ \mathrm{cm^2\ s^{-1}}$); however that value is (1) for the isotropic diffusion coefficient, a factor $\sim 3$ smaller than the parallel diffusivity, and (2) in \citet{girichidis2018cooler} the CR halo scale height adopted is smaller than the results of our cosmological simulations here, which in turn leads to a lower required $\kappa$ to reproduce the observations.}

It is worth noting that the diffusion coefficient is assumed to be constant in our simulations for simplicity, while in reality microphysical models indicate that diffusivity could be a highly complicated, nonlinear function of local plasma properties. Unfortunately, there does not yet exist a commonly accepted CR transport model. Recently, \citet{hopkins2020testing} explored varying CR transport coefficients in FIRE-2 cosmological zoom-in simulations, where the variations were motivated by different transport models available in the literature. \citet{hopkins2020testing} found that in different CR transport models that are allowed by observational constraints, the impact of CRs on CGM properties are qualitatively similar, but with quantitative differences in the strength of the CR effects.

\citet{hopkins2019but} studied $>30$ different zoom-in volumes: here we focus on a representative subset given in Table~\ref{tbl:sims}, but we have verified that all the qualitative behaviors here are robust across different halos of similar mass in the larger sample.

\section{Results}
\label{sect:results}

\subsection{CRs and CGM Pressure Support}
\label{sect:grad_P}

\begin{figure*}
\begin{centering} 
\includegraphics[width={\textwidth}]{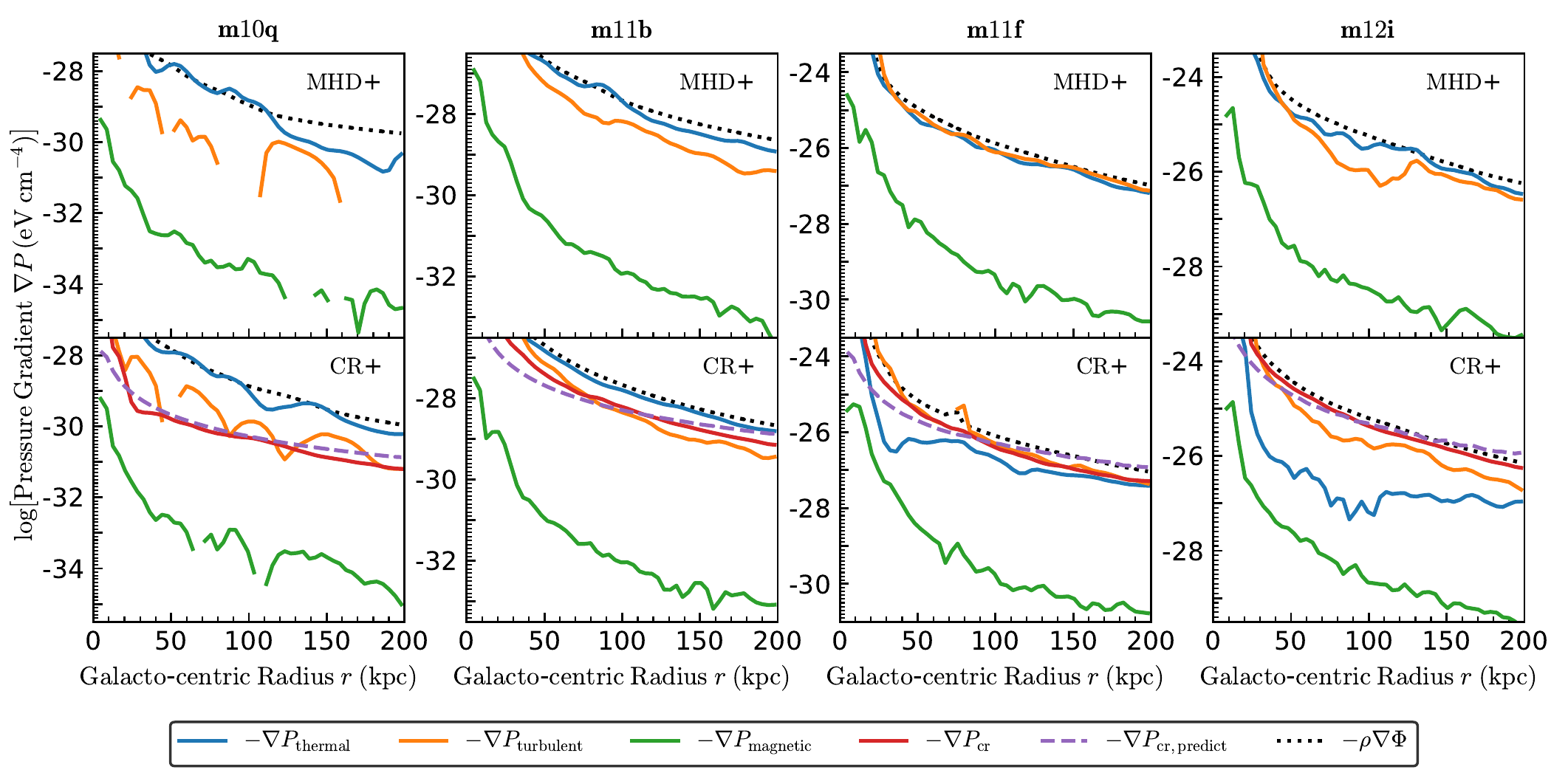}
    \end{centering}
    \vspace{-0.5cm}
        \caption{Radial profiles of the gas thermal pressure gradient $\nabla {P_\mathrm{thermal}} $ ({blue}), turbulent pressure gradient $\nabla P_\mathrm{turbulent}$ (orange), magnetic pressure gradient $\nabla P_\mathrm{magnetic}$ (green), CR pressure gradient $\nabla P_\mathrm{cr}$ ({red}), analytically predicted CR pressure gradient $\nabla P_\mathrm{cr,predict}$ (purple dashed, details presented in \S\ref{sec:discussion.scalings}) and gravitational force $\rho \nabla \Phi$ ({black} dotted) in the CGM, averaged in spherical shells as a function of galacto-centric radius $r$ at $z=0$ in our fully-cosmological simulations, where only the radial component of the gradient fields is taken into account. We compare simulations which neglect explicit CR dynamics/transport, i.e.\ our ``No CR'' or ``MHD+'' runs, and runs including our ``full'' CR physics with the constant diffusivity $\kappa_{\|}$ which best reproduces $\gamma$-ray observations (see \S~\ref{sect:methods} and \citealt{chan2019cosmic,hopkins2019but}). Different halos from Table~\ref{tbl:sims} are labeled. In dwarfs ({\bf m10q}, {\bf m11b}), CR pressure is always sub-dominant, consistent with weak effects of CRs on dwarfs seen in \citet{hopkins2019but}. In intermediate and MW-mass ({\bf m11f}, {\bf m12i}) systems, CR pressure can dominate thermal pressure and balance gravity, with relatively low {$P_\mathrm{thermal}$}, when CRs are included.
\label{fig:grad_P}\vspace{-0.2cm}}
\end{figure*}

Fig.~\ref{fig:grad_P} shows profiles of the gas (thermal) pressure and CR pressure gradients, for various representative halos at $z\sim0$ in our simulations, comparing our MHD+ and CR+ runs. The magnetic and turbulent pressure (defined as $P_\mathrm{magnetic}\equiv B^2/8\pi$ and $P_\mathrm{turbulent}\equiv \rho \delta v^2$ respectively) are also shown, but as discussed in more detail in \citep{hopkins2019but} they are negligible in the CGM (where turbulence is weak and the plasma $\beta \equiv P_{\rm thermal}/P_{\rm magnetic} \gg 1$ always) -- this is why we see (as \citealt{hopkins2019but}) only very small differences between our MHD+ runs and runs neglecting magnetic fields entirely (the ``Hydro+'' runs in \citealt{hopkins2019but}). We therefore will focus on the dominant pressure terms: thermal and CR. We compare these to the gravitational force per unit volume.

Fig.~\ref{fig:grad_P} shows that for lower-mass dwarfs (e.g.\ {\bf m10q} and {\bf m11b}), in both MHD+ and CR+ runs, thermal gas pressure is the leading term in balancing the gravitational force, and $P_{\rm thermal} / P_{\rm cr}\gg1$ in {\bf m10q} and $P_{\rm thermal} / P_{\rm cr}\sim 2$ in {\bf m11b}, with thermal pressure progressively more dominant at lower masses. Not surprisingly, in these cases we find CRs have relatively weak effects in the CGM, as \citet{hopkins2019but} also found for ISM and galaxy properties. This is true across the large ensemble of $\sim 20$ dwarf halos simulated in \citet{hopkins2019but}, so for brevity we simply focus on these representative cases. However, a qualitative change occurs for the CR+ run with intermediate-mass halos of {\bf m11f} through the MW-mass {\bf m12i} and {\bf m12m}: the CR pressure becomes dominant over thermal (and magnetic) pressure in the CGM, and balances gravity. Turbulent pressure can be comparable to CR pressure in {\bf m11f}, and slightly subdominant to, but still of the same order of magnitude to CR pressure in {\bf m12i}, suggesting it is not negligible, but is also not able to single-handedly provide the full pressure support needed in the CGM. Again, this is shown for a larger sample in \citet{hopkins2019but} -- here we focus on the representative cases shown, but also (by showing the pressure gradients instead of total pressure) demonstrate explicitly that the CR pressure gradient almost exactly balances gravity in the MW-mass systems (see \S\ref{apdx:pressure_m12} for radial pressure profiles of a larger sample of MW-mass halos).

\begin{figure*}
\begin{centering}
\includegraphics[width={\textwidth}]{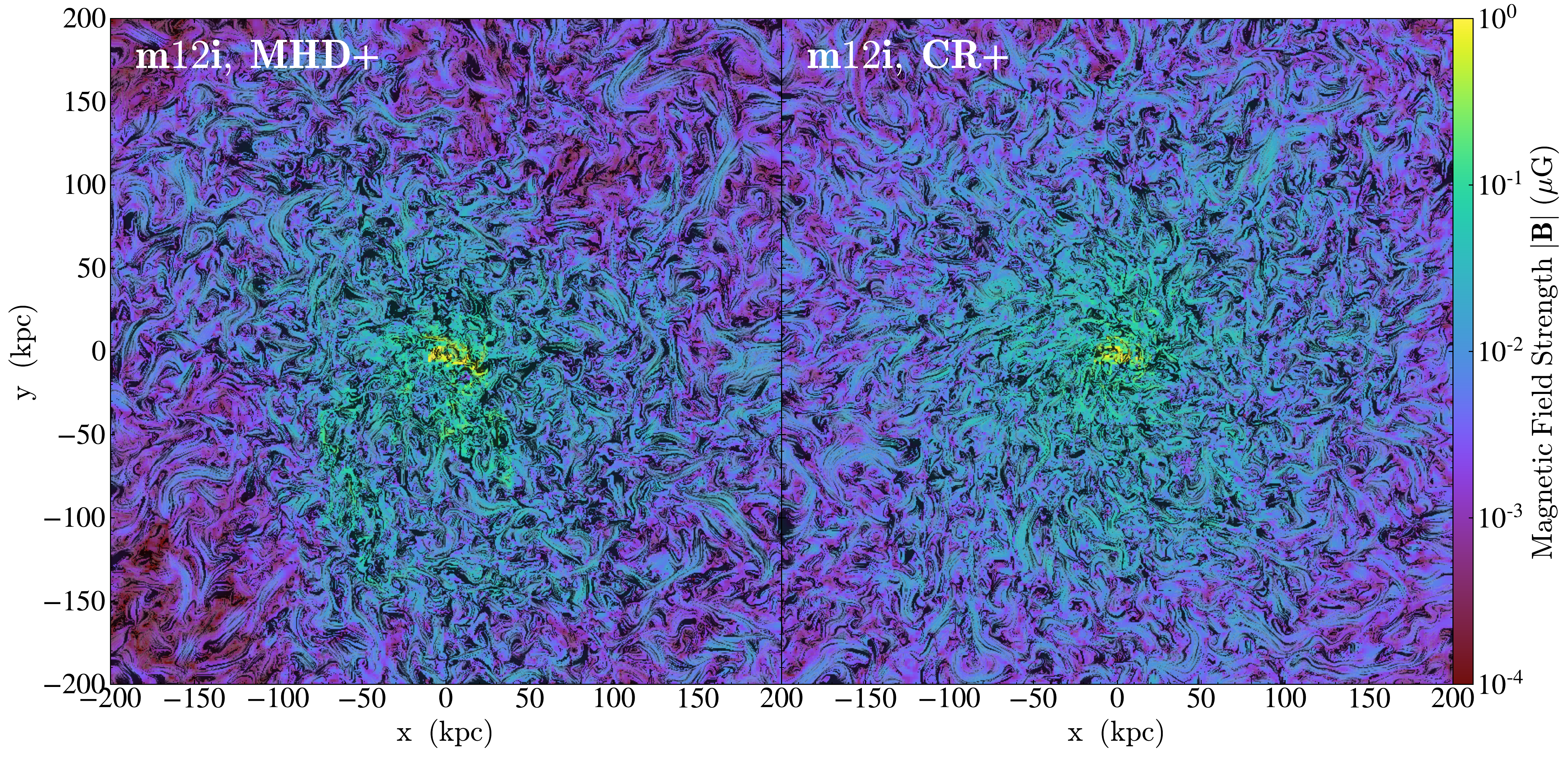}
    \end{centering}
    \vspace{-0.7cm}
        \caption{Magnetic field lines (black) and strength (color) from the {\bf m12i} MHD+ ({\it left}) and CR+ ({\it right}) runs at $z=0$, in a slice through the galaxy. The turbulence in the CGM, while weak (Fig.~\ref{fig:grad_P}), produces highly-tangled fields. The structure and amplitude (see also Fig.~\ref{fig:grad_P}) of magnetic fields in the CGM are similar with or without CRs -- i.e.\ CRs do not appear to dramatically ``open up'' field lines or suppress dynamo growth.
    \label{fig:Bfield}\vspace{-0.2cm}}
\end{figure*}

\subsection{CRs and CGM Magnetic Fields}
\label{sec:bfield}

Fig.~\ref{fig:Bfield} compares magnetic field structure and strength in the CGM for run {\bf m12i}. As shown in a more detailed study in  \citet{su2017stellar}, the fields are primarily amplified in the CGM by a combination of flux-frozen compression during collapse, stirring and transport of galactic fields via galactic winds, and the turbulent dynamo, giving rise to to an approximate $|\bm{B}|\propto n^{2/3}$ scaling with $\beta \gg 1$, which we also find (on average) here. We also clearly see that turbulence in the CGM has led to highly tangled fields, which explains why e.g.\ \citet{chan2019cosmic} and \citet{hopkins2019but} found that adopting isotropic diffusion or streaming with isotropically-averaged coefficient $\tilde{\kappa} \sim \kappa_{\|}/3$ gave similar results to the default fully-anisotropic transport here. The fields are also weak, consistent with our previous studies \citep{su2017feedback,su2019failure,hopkins2019but} and general expectations in the CGM, with $\beta \sim 100-10^{4}$ at $\sim 100-300\,$kpc. Most important, Figs.~\ref{fig:grad_P}-\ref{fig:Bfield} show that field strengths and morphologies are similar in MHD+ and CR+ runs even in {\bf m12i} (differences are even smaller in the dwarf runs): so CRs do not appear to strongly modify CGM magnetic fields.

\begin{figure}
\begin{centering}
\includegraphics[width={0.95\columnwidth}]{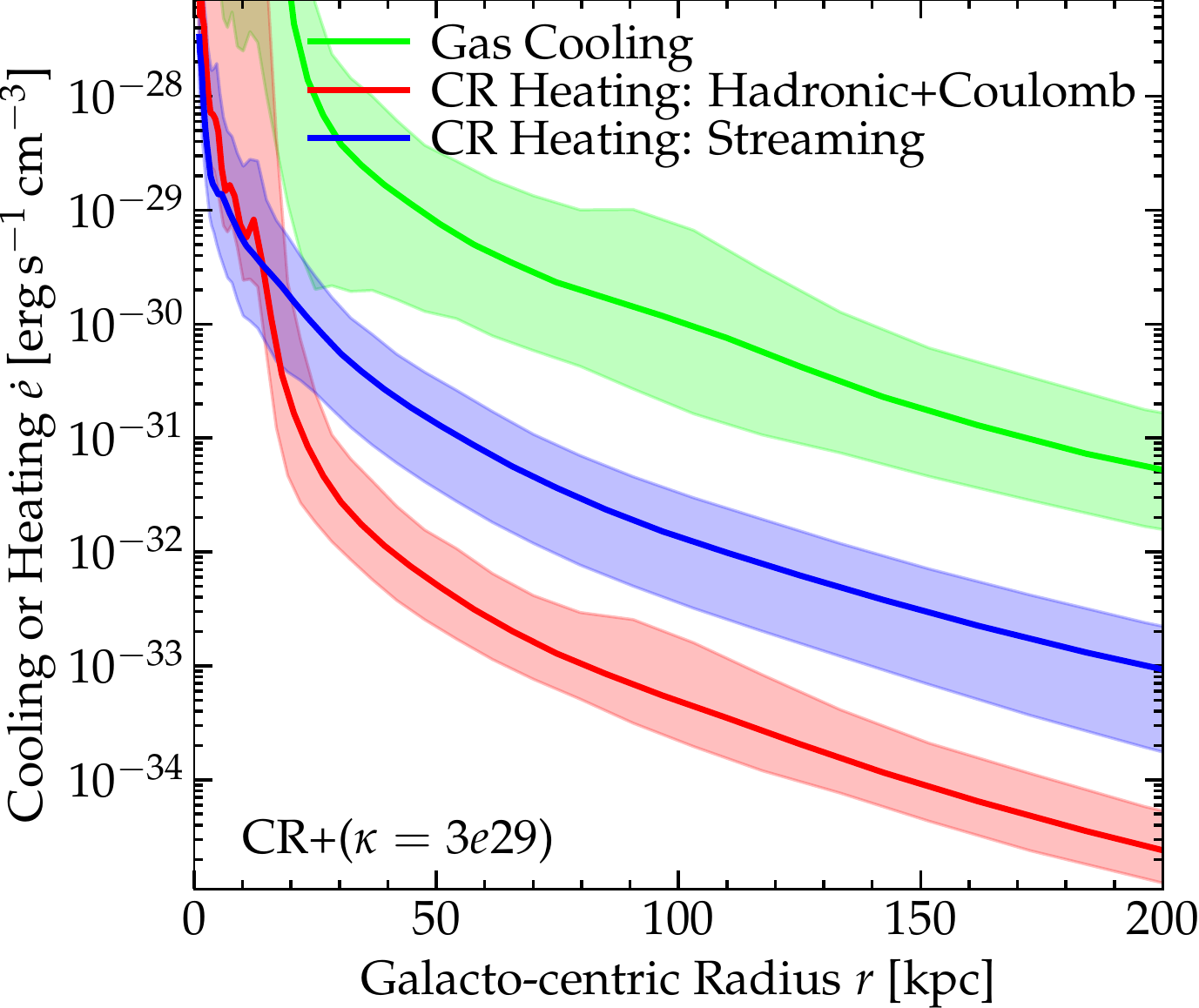} 
\end{centering}
    \vspace{-0.25cm}
        \caption{Radial profiles of gas heating and cooling rate, around {\bf m12i} in the CR+ run at $z=0$. Solid lines show the volume-averaged profile in spherical shells; shaded range shows the $5-95\%$ inclusion interval of all resolution elements at that radius. 
        We compare total gas cooling rate vs.\ heating rate from CRs via collisional (hadronic+Coulomb) and streaming losses. CR heating is weak compared to gas cooling in the CGM; the dominant term is the CR pressure (Fig.~\ref{fig:grad_P}), which explains why CGM gas in CR+ runs is {\it cooler}, not warmer, on average compared to MHD+ runs (Fig.~\ref{fig:profile_rhoT}). 
    \label{fig:profile.heating}\vspace{-0.1cm}}
\end{figure}

\subsection{CRs and CGM Heating}
\label{sec:heating}

CRs can in principle alter the CGM via their pressure support, or via heating the gas though either collisions (hadronic and Coulomb encounters which thermalize a fraction of the CR energy in each collision) or ``streaming losses'' (excitation of extremely high-frequency Alfv\'en waves as CRs stream via the gyro-resonant instability, which damp and thermalize their energy rapidly). Both of these are included in our CR treatment, so Fig.~\ref{fig:profile.heating} shows the profiles of thermal heating of the gas owing to each term, compared to the total gas cooling rate. Even in {\bf m12i} where the ratio of CR to thermal pressure is maximized, the CR heating is just a few percent of the total cooling rate; for less-massive halos the effect is even weaker \citep[see][]{hopkins2019but}. In MW and lower-mass halos, cooling is simply too efficient, with $\Lambda \gtrsim 10^{-22}\,{\rm erg\,s^{-1}\,cm^{3}}$ at the range of temperatures of interest. Thus, consistent with \citet{hopkins2019but}, we find CR heating is negligible for dwarf-through-MW mass halos at any radius.

\subsection{CGM Gas Phases}
\label{sect:phase}

\begin{figure*}
\begin{centering}
\includegraphics[width={0.4\columnwidth}]{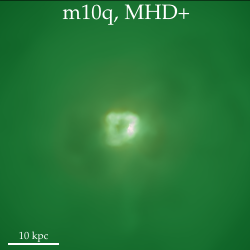}
\includegraphics[width={0.4\columnwidth}]{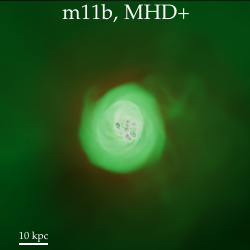}
\includegraphics[width={0.4\columnwidth}]{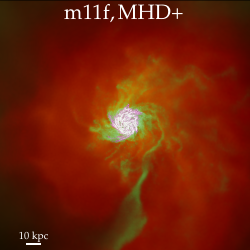}
\includegraphics[width={0.4\columnwidth}]{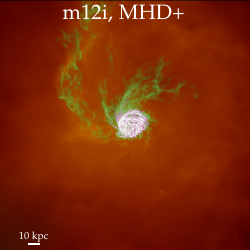}
\includegraphics[width={0.4\columnwidth}]{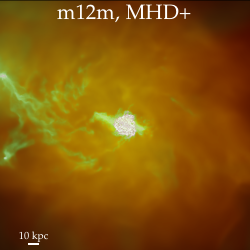} \\
\includegraphics[width={0.4\columnwidth}]{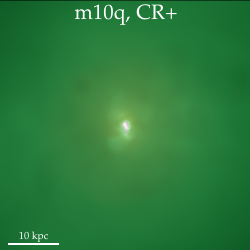} 
\includegraphics[width={0.4\columnwidth}]{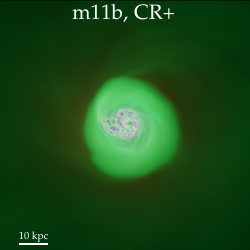}
\includegraphics[width={0.4\columnwidth}]{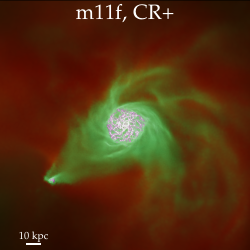}
\includegraphics[width={0.4\columnwidth}]{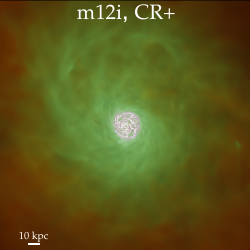}
\includegraphics[width={0.4\columnwidth}]{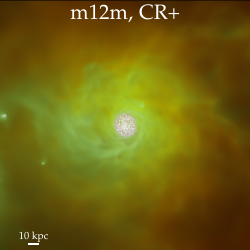} \\
    \end{centering}
    \vspace{-0.25cm}
        \caption{CGM gas morphologies of the galaxies from Table~\ref{tbl:sims} at $z=0$, with and without CRs.
         Images are three-band volume renders of gas showing {warm/}hot ($T\gg 10^{5}\,$K; {\it red}), {cool} ($T\sim 10^{4}-10^{5}$\,K; {\it green}), and cold neutral ($T\ll 10^{4}\,$K; {\it magenta/white}) phases. Boxes extend to $\sim 1/2$ of the virial radius, with scale bars labeled. 
         Systematic differences are minor at dwarf masses ({\bf m10q}, {\bf m11b}); but in low-$z$, MW-mass systems where CRs dominate the pressure in CR+ runs (Fig.~\ref{fig:grad_P}), we see gas shift from hot to warm phases. 
    \label{fig:morph}\vspace{-0.5cm}}
\end{figure*}

\begin{figure}
\begin{centering}
\includegraphics[width={\columnwidth}]{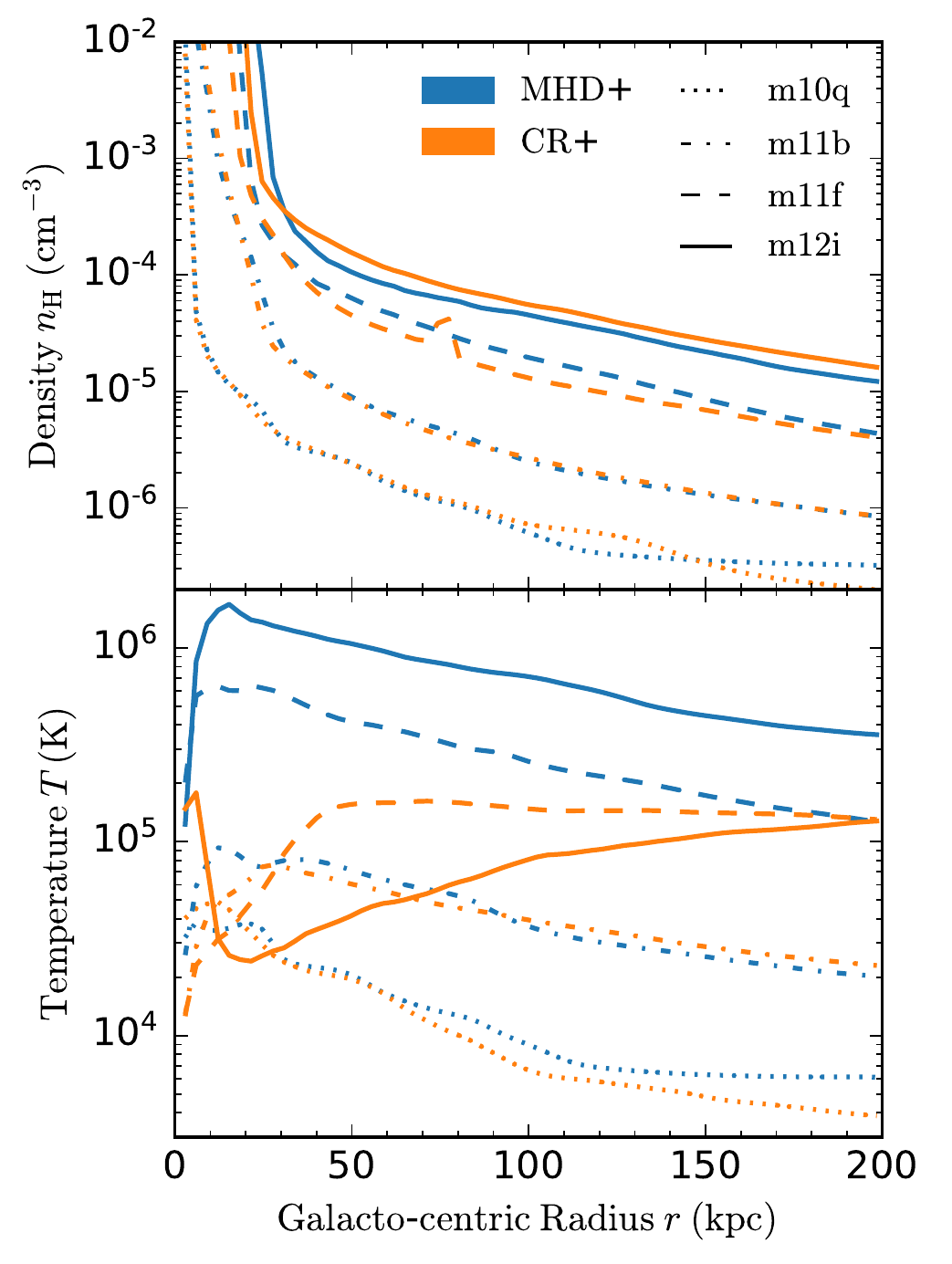}
\vspace{-0.6cm}
\end{centering}
\caption{Radial profile of gas mass density ({\it top}) and temperature ({\it bottom}) {weighted by volume} for halos from Fig.~\ref{fig:grad_P}, in MHD+ and CR+ runs. Plotted values are volume-averaged in spherical shells at galacto-centric radius $r$. Again, effects of CRs in dwarfs are minor. In massive halos, CR+ runs do not strongly modify the global volume-averaged gas density profiles or total halo baryonic mass, but do strongly alter the gas temperature structure, with significantly lower median gas temperatures evident from $\sim 10-200\,$kpc.
\label{fig:profile_rhoT}\vspace{-0.2cm}}
\end{figure}

Fig.~\ref{fig:morph} shows a volume render of the CGM gas, highlighting different gas phases. At low (dwarf) halo masses {\bf m10b} and {\bf m11b}, there is no systematic difference between MHD+ and CR+ runs -- the same is seen in similar volume-renders {\it within} the galaxies in \citet{hopkins2019but}. However, as halos approach MW mass ($M_{\rm halo} \gg 10^{11}\,M_{\odot}$, i.e. {\bf m11f} and {\bf m12i}), where we saw CR pressure dominate in the CGM in Fig.~\ref{fig:grad_P}, we see a significant increase in the prominence and volume-filling-factor of cool and warm gas in the CR+ runs. In contrast, the MHD+ runs are dominated by hot gas, with warm/cool gas in the CGM restricted to filamentary structures with small volume-filling factor. Fig.~\ref{fig:profile_rhoT} shows this quantitatively, plotting the gas density and temperature as a function of galacto-centric radii. The gas density profiles are similar between CR+ and MHD+ runs at all halo masses, but the temperature profiles differ dramatically in the CR+ runs of MW-mass halos. In e.g.\ {\bf m12i}, the median temperature in the MHD+ run peaks at $\sim 10^6 \, \mathrm{K}$ at $r\sim 20\,$kpc then slowly decays to $\sim 3 \times 10^5 \, \mathrm{K}$ at $r \sim 200 \, \mathrm{kpc}$, while in the CR+ run it rises (outside the disk at $r\lesssim 10\,$kpc) from $\sim 3\times10^{4}\,$K at $r\sim 20\,$kpc to $\sim 10^{5}\,$K at $r\sim 200\,$kpc. Halo {\bf m11f}, being intermediate-mass, shows the same effect of CRs, but less dramatically. It is worth noting that although the total gas density profiles in Fig. \ref{fig:profile_rhoT}, i.e. the total masses at any given radii, are similar between the MHD+ and CR+ runs, the detailed gas density and temperature distributions are significantly different. We discuss these phase structure differences in more detail in the following sections.

\begin{figure}
\begin{centering}
\includegraphics[width={0.95\columnwidth}]{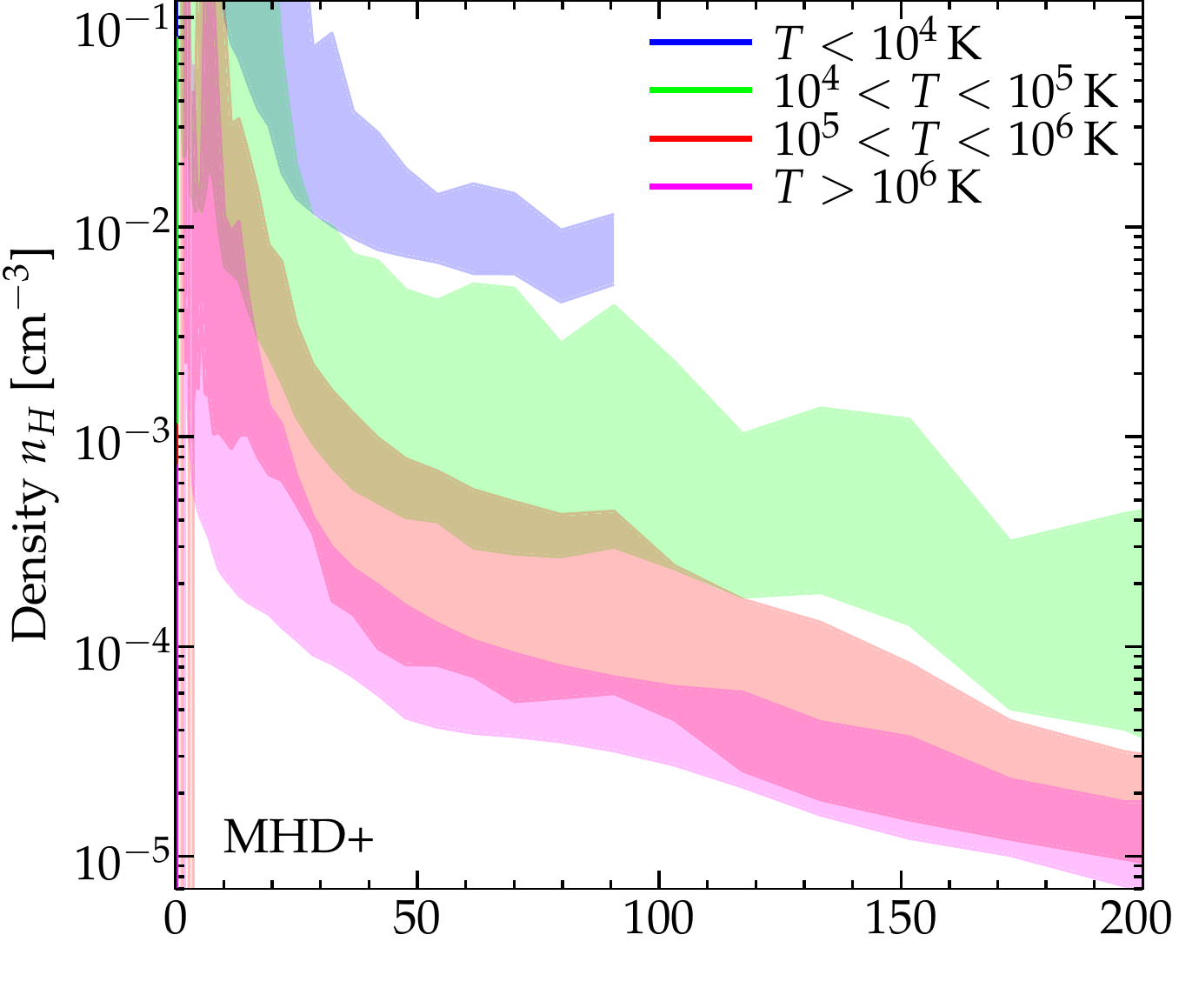} \\
\vspace{-0.35cm}
\includegraphics[width={0.95\columnwidth}]{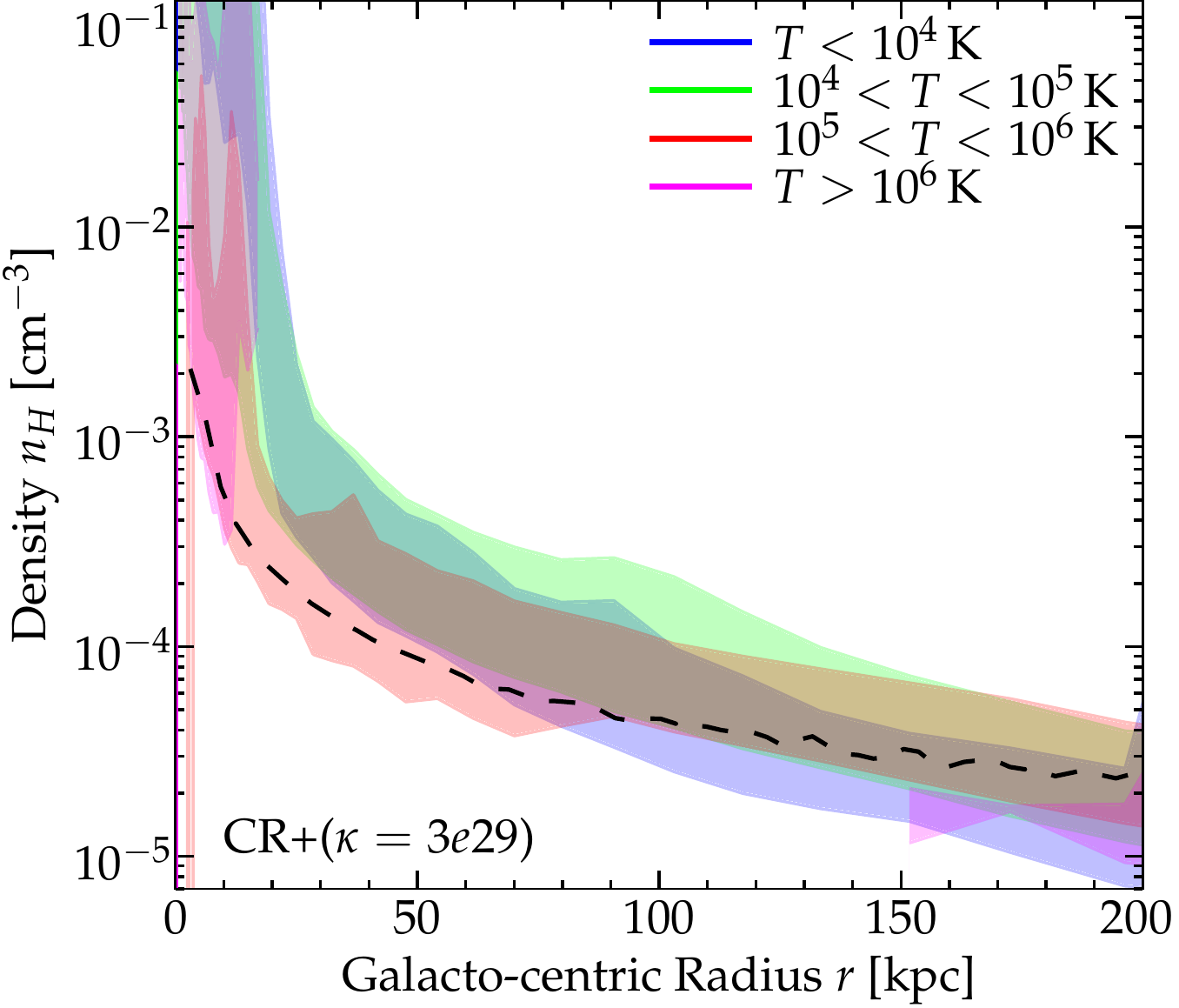} \\
    \end{centering}
    \vspace{-0.25cm}
        \caption{Gas density profiles {\it in different phases} for {\bf m12i} at $z=0$, comparing MHD+ and CR+ runs. 
        We specifically compare the $5-95\%$ range of {\it local} gas densities $n_{\rm H}$, for all resolution (gas mass) elements at a given radius and within a given temperature range (colors, labeled). Note that because this is the {\it local} density for only gas with a given $T$ (i.e., the gas mass in some phase divided by only the volume occupied by the gas in that phase, not by the entire volume), the value of $n_{\rm H}$ here does not correspond to the mass fraction in that phase (e.g.\ cold gas can have large $n_{\rm H}$, but low volume-filling-factor, hence a low mass fraction). 
         In MHD+ runs, gas at low-$T$ has higher {local} $n_{\rm H}$ -- i.e.\ gas at the same radius is in approximate local {\it thermal} pressure equilibrium. 
         In CR+ runs, all phases have similar local $n_{\rm H}$, required for CR pressure to balance gravitational force (Fig.~\ref{fig:grad_P}), independent of $T$ to first order. An analytic density profile assuming pure CR pressure support is plotted by the black dashed curve. This simple model is in good agreement with the actual gas density profiles in the simulations for all phases at the larger radii where the CR pressure is dominant (see \S\ref{sec:discussion.scalings} for the analytic model).
    \label{fig:profile.density}\vspace{-0.2cm}}
\end{figure}

\begin{figure*}
\begin{centering}
\includegraphics[width={\textwidth}]{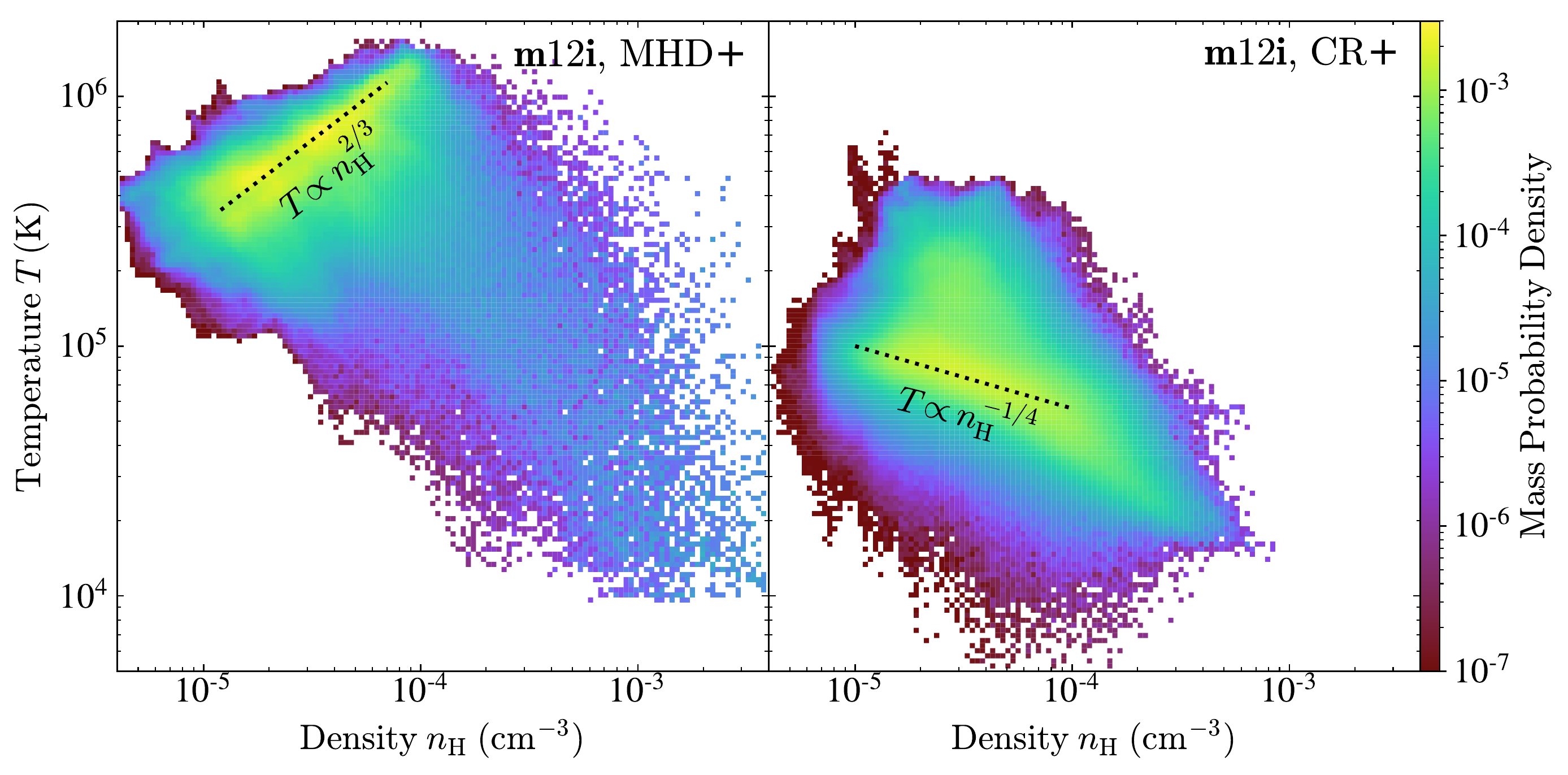}
\vspace{-0.25cm}
\end{centering}
\caption{Density-temperature phase diagrams of gas in MHD+ and CR+ runs of {\bf m12i} at $z=0$. We show gas in the CGM with $50\ \mathrm{kpc} < r < 200 \, \mathrm{kpc}$. Colors show the probability density weighted by gas mass. The CR+ run is cooler, but also follows a different trend: while the probability ``ridgeline'' in MHD+ follows an approximate adiabat ($T\propto n^{2/3}$), the corresponding line in CR+ more closely follows the expectation from photo-ionization equilibrium ($T \propto n^{-1/4}$, e.g., \citealt{stern2018does}).
\label{fig:phase_dens_temp}\vspace{-0.5cm}}
\end{figure*}

Figs.~\ref{fig:profile.density}-\ref{fig:phase_dens_temp} examine the gas phases in more detail, in run {\bf m12i} where the impact of CRs is most apparent. Fig.~\ref{fig:profile.density} shows the distribution of local densities $n$ for gas at different specific temperatures $T$ and radii $r$: this allows us to directly compare the typical thermal pressure of different phases as $\sim n\,k_{B}\,T$ at each $r$. In the MHD+ runs, we see roughly $n \propto T^{-1}$ at each $r$, i.e.\ $P_{\rm thermal}(r)$ is constant independent of $T$, so the gas phases are in {\it local} thermal pressure equilibrium with each other, in addition to ``global'' thermal pressure equilibrium with gravity (seen in Fig.~\ref{fig:grad_P}). But in the CR+ run, the different phases all reside at approximately the same density (at a given $r$), independent of $T$, so $P_{\rm thermal}(r) \propto T$ at a given $r$, meaning that the cold(er) phases are ``under-pressurized'' relative to hot phases at the same radius, in addition to the total thermal pressure being globally below that needed for virial equilibrium (Fig.~\ref{fig:grad_P}).

Fig.~\ref{fig:phase_dens_temp} shows the complementary $n-T$ diagram over a broad range of $r$ in the CGM. For the MHD+ run, most of the gas is in the hot phase above $10^5 \, \mathrm{K}$. There is a ``ridge'' of high probability density which directly corresponds to the median $n(r)$ and $T(r)$ radial trend in Fig.~\ref{fig:profile_rhoT}, i.e.\ position along the ridge is a progression in $r$, along a simple adiabat with $T(r) \propto n(r)^{2/3}$. In the CR+ run, the temperatures are cooler, and there is an analogous ridgeline tracing the radial profile, but here temperature anti-correlates with density, approximately following $T(r) \propto n(r)^{-0.25}$ as expected for diffuse gas in photo-ionization equilibrium with the UV background at $z\sim 0$ (see e.g., \citealt{stern2018does}). 

Our findings that CRs can provide significant non-thermal pressure support in the CGM are qualitatively similar to the conclusions of \citet{salem2016role}. However, we find in this work that the CR effects can be stronger. In particular, for our fiducial assumptions, we find that the CR pressure can dominate the thermal pressure by more than one order of magnitude in the entire CGM. On the other hand, in \citet{salem2016role} the volumes primarily supported by thermal pressure vs. CR pressure are comparable, with the CR pressure-supported cool gas limited to swath-like regions. As a result, there is more cool gas in the CGM of our Milky Way-mass simulations. This difference could be primarily due to our treatment of anisotropic CR diffusion and streaming with non-adiabatic CR energy losses, and a higher CR diffusion coefficient adopted in our simulations. With this CR treatment, the $\gamma$-ray emission of the whole galaxies matches observations very well over a wide range of galaxy masses \citep{chan2019cosmic}. \citet{salem2016role} instead assumes isotropic diffusion with a lower CR diffusion coefficient, neglecting CR streaming and hadronic losses. They only considered the $\gamma$-ray emission in the CGM, but did not calculate or compare the $\gamma$-ray emission of the whole galaxies with observations.

\subsection{Column Densities of Different Ions in the CGM}
\label{sect:column}

Given the difference in gas phases, it is natural to expect differences in observed column densities of different temperature-sensitive ions in the CGM. We post-process the simulations with {\small Trident} \citep{hummels2017trident} to calculate ion-number densities, using the simulation density, temperature, and metal abundances for each species, accounting for collisional and photo-ionization (with self-shielding included, where the ionization depth equals the local Jeans length with a maximum depth of $100\,\mathrm{pc}$ when the self-shielding ion table is generated) from the UV background, and then integrate these alone sightlines with an ensemble of random viewing angles at each impact parameter and sampling each simulation snapshot with spacing of $\sim 10\,$Myr over the redshift range of interest. We then measure the median and distribution of column densities for each ion over the ensemble of sightlines. A detailed description of this methodology, as well as the dependence of predicted column densities on e.g.\ numerical resolution, numerical methods for e.g.\ hydrodynamics and metal diffusion, halo mass, redshift, and other (non-CR) physics will be the subject of a companion paper (Hummels et al., in prep) -- our focus here is exclusively on the systematic effect of CRs on the predicted columns. 

The choice of UV background model is important for photoionization modeling. 
As shown in, for example, \citet{faucher2020cosmic} the \citet{faucher2009new} (FG09) UV background more accurately reproduces observations of the low-redshift ($z\lesssim 0.5$) Ly$\alpha$ forest compared to \citet{haardt2012radiative} (HM12), which under-predicts observationally-inferred \HI photo-ionization rates by a factor $\sim 2$. 
However, FG09 did not include a detailed treatment of the contribution of obscured and non-obscured AGN to the UV background. 
As a result, the FG09 background likely under-estimates the low-redshift UV at energies higher than 4 Ry.\footnote{The newer \cite{faucher2020cosmic} UV background model does include an improved treatment of obscured and non-obscured AGN, but {\small Trident} photoionization tables for this new model are not yet available.} We therefore use a combination of the FG09 and HM12 background models for our ionic predictions. Specifically, we use FG09 for \HI, \MgII, \SiIV and HM12 for \NV, \OVI, \NeVIII. We show the effects of different UV background assumptions in Appendix~\ref{apdx:hm_vs_fg}.

\begin{figure*}
\begin{centering}
\includegraphics[width={\textwidth}]{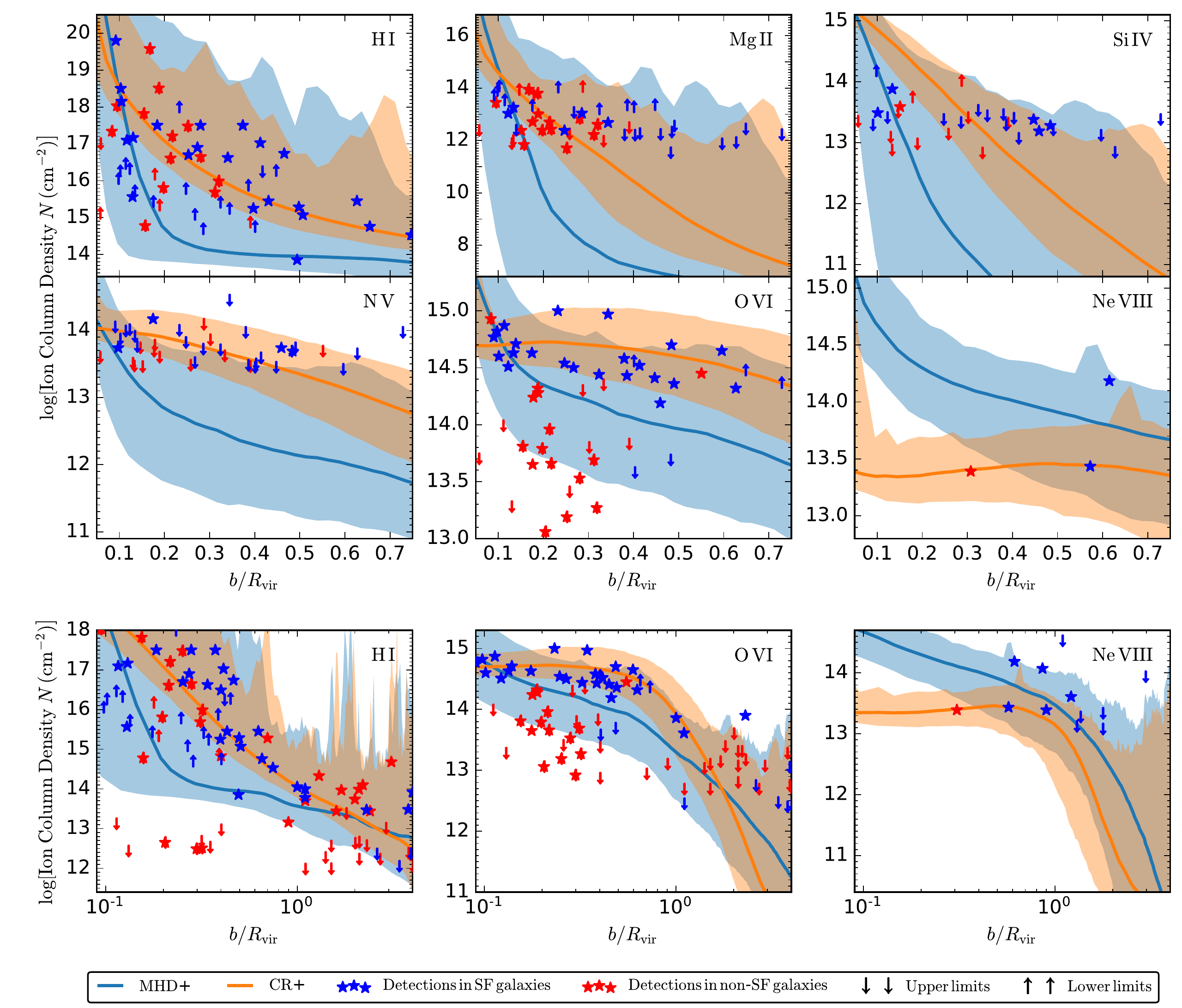}
\end{centering}\vspace{-0.25cm}
\caption{Column density profiles of various ions, as a function of impact parameter normalized to virial radius $b/R_\mathrm{vir}$. We compare MHD+ and CR+ for halo {\bf m12i} whose virial radius is $R_\mathrm{vir}^\mathrm{m12i} = 270\,\mathrm{kpc}$. Lines are (linear) averages over an ensemble of sightlines and times, and shaded range shows the full range across $800^2$ sightlines. Column density profiles are plotted on a linear (logarithmic) scale up to $\sim 0.8R_\mathrm{vir}$ or $200\,\mathrm{kpc}$ in {\bf m12i} ($ 4 R_\mathrm{vir}$ or $1100\,\mathrm{kpc}$ in {\bf m12i}) in the top two rows (bottom row). We compare observed columns along each slightline compiled from \citet{werk2013cos,johnson2015possible,j_xavier_prochaska_2017_1045480,prochaska2017cos,burchett2018cos,zahedy2019characterizing}, around star-forming and non-star-forming galaxies, with both detections and upper/lower limits (stars and arrows, respectively). To match the observations {and simulations}, we {only show observations in host galaxies with stellar masses $M_*=10^{9.5-11.5}\,M_\odot$, and} sample the simulations over redshifts $z=0.15-0.35$ except for \NeVIII which samples $z=0.6-0.8$ as observed. 
As explained in \S \ref{sect:column}, the FG09 UV background model is assumed for lower-energy ions (\HI, \MgII, \SiIV) and the HM12 background is used for higher-energy ions (\NV, \OVI, \NeVIII). CRs shift gas from hot to warm/cool phases, decreasing high ions and enhancing low/mid ions. 
\label{fig:column_m12i}\vspace{-0.5cm}}
\end{figure*}

Fig.~\ref{fig:column_m12i} compares the predicted ion column densities to observations. In the top two rows, predicted column density profiles are plotted as a function of impact parameter (normalized to virial radius which is $270\,\mathrm{kpc}$ for {\bf m12i}) up to $\sim 0.8 R_\mathrm{vir}$ (corresponding to $200\,\mathrm{kpc}$ in {\bf m12i} halo), with observations of \HI from \citet{prochaska2017cos,j_xavier_prochaska_2017_1045480} and \MgII, \SiIV, \NV, \OVI from \citet{werk2013cos} at $z\sim 0.15-0.35$; and \NeVIII from \citet{burchett2018cos} at $z\sim 0.6-0.8$. On the bottom row, detections and upper limits of \HI, \OVI and \NeVIII are plotted on a logarithmic scale up to a longer distance of $4 R_\mathrm{vir}$ ($1100\,\mathrm{kpc}$ in {\bf m12i} halo), with \HI and \OVI observations at large radii from \citet{johnson2015possible} included. We discuss each in turn:

\subsubsection{Low Ions}

Median \HI columns in our CR+ run at $r\gtrsim 0.15 R_\mathrm{vir}$ are higher than those in the MHD+ run by factors $\sim 10-100$. For MHD+, the median value of \HI sharply declines and reached $10^{14.5}\,\mathrm{cm^{-2}}$ at $0.2R_\mathrm{vir}$, while in CR+ median declines gently and reaches the same column at $r\sim 0.7 R_\mathrm{vir}$.

Note that the {\it maximum} \HI columns at any radius are similar or even higher in MHD+; in fact, the entire range of columns in the CR+ run lies within the range seen in MHD+, which also extends to much lower minimum \HI columns. Also, the total \HI mass in the CGM is similar for both MHD+ and CR+ runs. The difference in median and minimum columns can be explained by the morphology of the cool phase: in CR+ the cool-phase is volume-filling, while it is confined to dense filaments with small volume-filling-factor in MHD+. Thus, the median \HI column across random sightlines is much higher in the CR+ run, and the dispersion or sightline-to-sightline variance is much smaller. The observed \HI lower limits ($\gtrsim 10^{15} \, \mathrm{cm^2}$) at $\sim 0.1-0.4 R_\mathrm{vir}$ and median detections at $\sim 0.2-0.7 R_\mathrm{vir}$ suggest typical \HI columns significantly larger than the median in the MHD run, and comparable to the predictions of the CR+ run, especially if we restrict to observations around star-forming galaxies (which {\bf m12i} is).

Qualitatively similar behavior is evident in \MgII, but given that most observations at $r\gtrsim 50\,\mathrm{kpc}\,(0.2 R_\mathrm{vir})$ are upper limits (and both CR+ and MHD+ runs produce similar maximal columns covering the range observed), the observations do not yet strongly favor one model or the other.

\subsubsection{Mid Ions}

\SiIV columns show similar behavior to \MgII and \HI, with a slower drop in CR+ runs (higher median and minimum columns at large $r$). Interestingly, neither MHD+ nor CR+ well match observational constrains: the detections and upper limits seem to suggest a flatter \SiIV profile, compared with which the CR+ predicted column density is too high in the inner halo ($\lesssim 0.2$) while the MHD+ predicted is too low in the outer halo ($\gtrsim 0.2 R_\mathrm{vir}$).

For \NV and \OVI at higher energy levels, we again see columns of \NV and \OVI fall more rapidly with $r$ in MHD+ runs compared to CR+ runs. In \NV, the median columns are systematically higher in the CR+ run by one order of magnitude at all $r\gtrsim 0.2 R_\mathrm{vir}$. The CR+ run better overlaps with the detections and upper limits, though if these upper limits reveal much lower columns it would favor the MHD+ run. For \OVI, columns are enhanced by a factor $\sim 2$ at $r\gtrsim 0.2 R_\mathrm{vir}$ in CR+ compared to MHD+, and the profiles are relatively flat, with characteristic columns $\sim 10^{14.5}\,{\rm cm^{-2}}$ at $\sim 0.55 R_\mathrm{vir}$ which corresponds to $150\,\mathrm{kpc}$ in {\bf m12i}. The \OVI profile in CR+ at a larger scale, as shown in the mid-bottom panel, remains flat with a significantly higher normalization at $r \lesssim 0.6 R_\mathrm{vir}$, and features a sharper cutoff at $\sim R_\mathrm{vir}$ than that in MHD+, which is more consistent with the detections and limits.

Indeed, the predicted difference in \OVI between CR+ and MHD+ runs, falling within one order of magnitude, is less dramatic if compared to a difference of $2-3$ orders of magnitude in \HI, in terms of the absolute numbers. However, we emphasize that the enhancement of \OVI in CR+ runs is a qualitative change, since it could potentially solve the open problem of \OVI as discussed in a substantial amount of recent papers (e.g., \citealt{faerman2017massive,oppenheimer2017flickering,mathews2017circumgalactic,stern2018does,stern2019cooling}). Although $N_\mathrm{OVI}\sim10^{14}\,\mathrm{cm^{-2}}$ can be reached with the assumption of a cooling quasi-static halo, many previous attempts struggle to get $N_\mathrm{OVI}\sim10^{14.5}\,\mathrm{cm^{-2}}$, by assuming either (1) some very strong heating mechanism which might overheat the low ions, (2) an extended \OVI halo beyond $R_\mathrm{vir}$ which is inconsistent with the observed sharp cutoff, or (3) the \OVI is photo-ionized and thus has such a low thermal pressure that some other form of pressure support is required. In our CR+ runs, as discussed later, it is the last case, where the low-pressure, photo-ionized \OVI gas is supported by the CR pressure.

\subsubsection{High Ions}

Columns in \NeVIII, on the other hand, are reduced in CR+ runs relative to MHD+ runs, consistent with an overall cooler CGM. The differences are relatively small beyond $\sim 0.7 R_\mathrm{vir}$, but while the median \NeVIII profile in CR+ remains flat with columns $10^{13.5} \, \mathrm{cm^{-2}}$, in MHD+ it rises monotonically towards smaller $r$ as the medium CGM temperature also increases (Fig.~\ref{fig:profile_rhoT}), to $\sim 10^{14.3}\,$ at $r\sim 0.2 R_\mathrm{vir}$. With the FG09 UV background, \NeVIII columns in CR+ runs are lower by another factor $\sim 2$. Observationally, two detections appear more consistent with our MHD+ runs, while the other two favor the CR+ runs. This suggest more hot gas might be needed in halos around $L_{\ast}$ star-forming galaxies, but low number of detections makes drawing a strong conclusion difficult.

\subsection{Collisional vs.\ Photo-Ionization}
\label{sect:ionization}

\begin{figure*}
\begin{centering}
\includegraphics[width={\textwidth}]{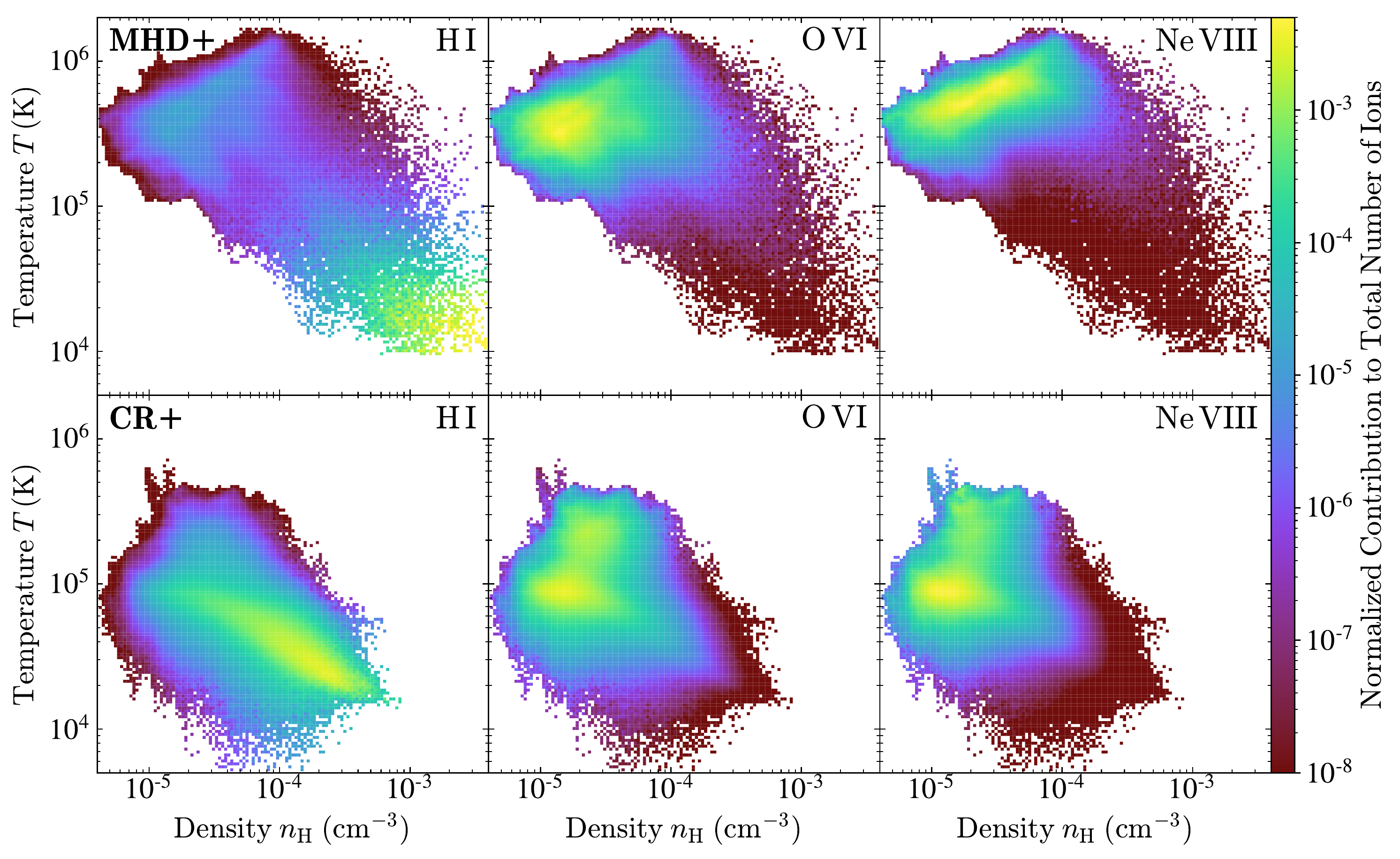}
\end{centering}
\vspace{-0.5cm}
\caption{Density-temperature diagrams of {\bf m12i} MHD+ ({\it top}) and CR+ ({\it bottom}) for CGM ($50\,{\rm kpc}<r<200\,{\rm kpc}$) gas at $z=0$, in the style of Fig.~\ref{fig:phase_dens_temp}. The only difference with Fig.~\ref{fig:phase_dens_temp} is that the probability density (colors) are now weighted by the number of ions of a given species (columns, as labeled), rather than the gas mass. While \HI traces the coolest, most dense CGM gas in both cases, in the CR+ runs we see the dominant contribution to \OVI and \NeVIII shift from the collisional peak at $T\gtrsim ({\rm few}) \times 10^{5}\,$K to the photo-ionization peak at lower temperatures.
\label{fig:phase_ion}\vspace{-0.5cm}}
\end{figure*}

Fig.~\ref{fig:phase_ion} shows the density-temperature diagram of CGM gas (as Fig.~\ref{fig:phase_dens_temp}), weighted by the number of ions in \HI, \OVI, and \NeVIII, for {\bf m12i} at $z\sim0$. In MHD+, most \HI comes from the ``tail'' (containing little mass) of the population with $T\sim 1-2\times10^{4}\,$K and $n\sim 1-5 \times 10^{-3}\,{\rm cm^{-3}}$, corresponding to the coolest and most dense filaments in the CGM, which are resistant to both collisional ionization (due to their low temperature) or photo-ionization (due to their high density). But in the CR+ run, \HI is mainly contributed by the diffuse and volume-filling gas at a temperature range from $10^4-10^5 \, \mathrm{K}$. In spite of the low (but non-zero) neutral fraction in this temperature range, the total amount of gas falling in this range is much larger than that at higher densities and lower temperatures, so it dominates the \HI columns. 

For MHD+, the \OVI is distributed along two ``strips'' on the plot: one is horizontal at $T\sim 3\times10^5 \, \mathrm{K}$, which is the temperature at which the \OVI ion fraction peaks due to collisional ionization; the other follows a positive slope along which the decrease of the \OVI ion fraction due to higher $T$ is compensated by larger amounts of gas along the ``ridgeline'' where most of the gas resides in Fig.~\ref{fig:phase_dens_temp}. In either case, most \OVI is contributed by collisional ionization from the warm halo gas with temperature $\gtrsim 10^5 \, \mathrm{K}$. For CR+, there is far less gas at high temperatures, so we see some of the \OVI coming from collisional ionization with $T\sim2-5\times 10^{5}\,$K, while a comparable but somewhat larger fraction comes from photo-ionized gas with $T\sim 0.7-1.2\times10^{5}\,$K and $n\sim 0.7-5\times10^{-5}\,{\rm cm^{-3}}$. We note that this is where the choice of UV background makes some difference: the same qualitative differences appear with the FG09 background, but the photo-ionization of gas with $T\sim 10^{5}\,$K and $n\sim 10^{-5}\,{\rm cm^{-3}}$ is somewhat less efficient, giving rise to overall less \OVI and moving the balance between collisional and photo-ionization somewhat more in favor of collisional ionization.

For \NeVIII, the same qualitative effect is even more pronounced: almost all the \NeVIII in MHD+ comes from the ``ridgeline'' of hot gas from Fig.~\ref{fig:phase_dens_temp} with $T\gtrsim 4\times10^{5}$\,K where it can be collisionally ionized, while in CR+ there is so little gas at these high temperatures that photo-ionization of gas with $T\sim 10^{5}\,$K, $n\sim10^{-5}\,{\rm cm^{-3}}$ dominates \NeVIII. Photo-ionization is less efficient at ionizing the higher-energy \NeVIII, so this produces overall less \NeVIII column compared to MHD+ case, as seen in Fig. \ref{fig:phase_dens_temp}.

\subsection{Dependence on Halo Mass and Redshift}

\begin{figure*}
\begin{centering}
\includegraphics[width={\textwidth}]{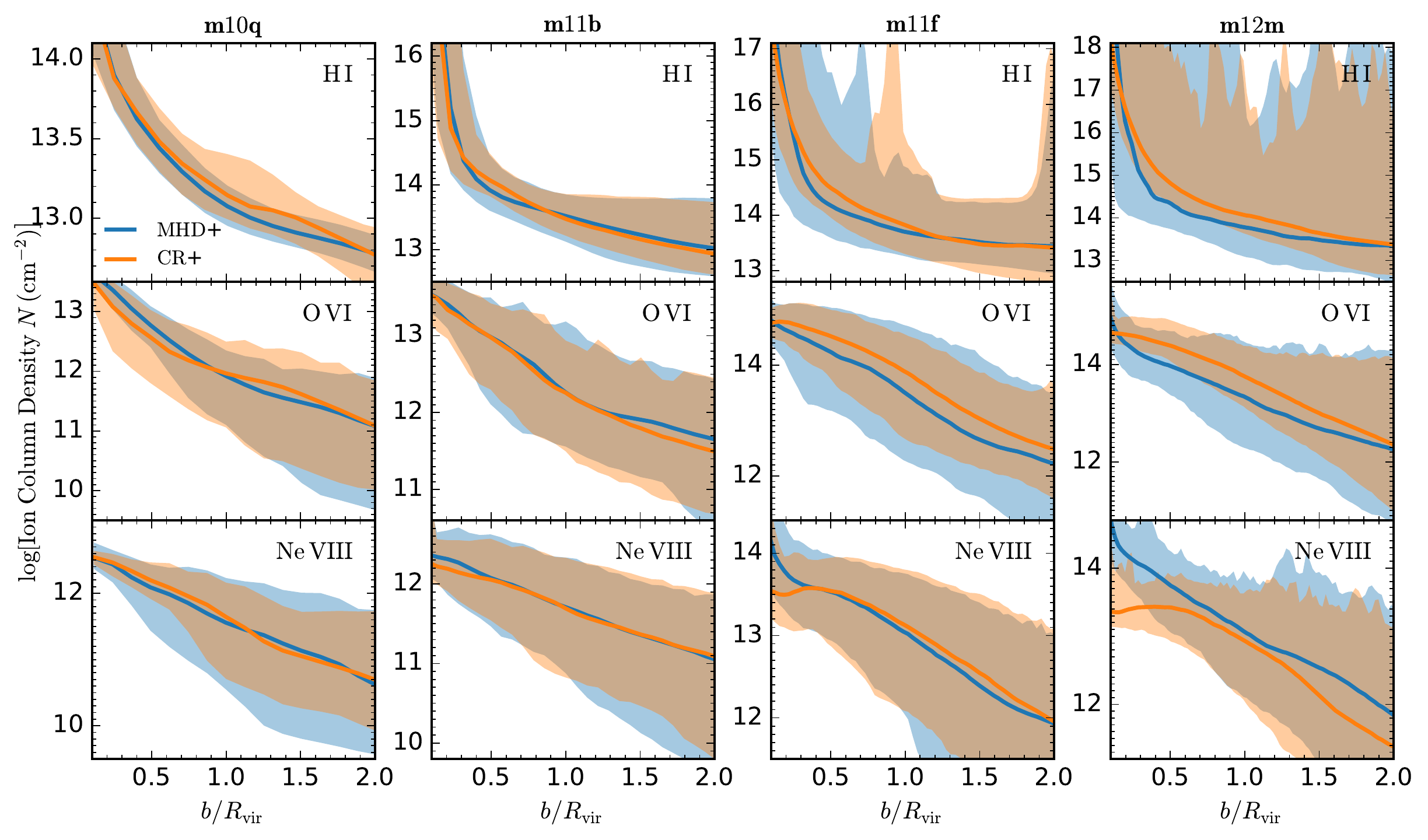}
\end{centering}
\vspace{-0.6cm}
\caption{Column density profiles, as Fig.~\ref{fig:column_m12i}, for different halos from Table~\ref{tbl:sims}. Once again, CRs have negligible effects in dwarfs ({\bf m10q}, {\bf m11b} here). Other MW-mass halos ({\bf m12m}) at $z\sim0-1$ exhibit the same behavior as {\bf m12i} (Fig.~\ref{fig:column_m12i}). At intermediate masses ($10^{11}\,M_{\odot} \lesssim M_{\rm halo} \lesssim 10^{12}\,M_{\odot}$, {\bf m11f}), where the halo has lower virial temperature (below that needed to excite \NeVIII) and more cold gas even without CRs, the low ions (\HI) are similar, but \OVI is enhanced at large radii and \NeVIII is suppressed at small radii, though the effect is modest. 
\label{fig:column_halos}\vspace{-0.5cm}}
\end{figure*}

Fig.~\ref{fig:column_halos} extends our comparison from Fig.~\ref{fig:column_m12i} to halos of different mass, focusing on \HI, \OVI, and \NeVIII as representative ions. As expected from Figs.~\ref{fig:grad_P}-\ref{fig:profile_rhoT}, for dwarfs ({\bf m10q}, {\bf m11b}), there is no significant difference between MHD+ and CR+ runs. For the other MW-mass halo {\bf m12m}, despite having a very different halo and star formation history and substantially different galaxy structure \citep[see][]{hopkins:fire2.methods}, the effects of CRs are very similar to {\bf m12i} but the magnitude of difference is less pronounced.. For the intermediate-mass {\bf m11f}, the differences in \HI and \OVI between CR+ and MHD+ runs are qualitatively similar (but smaller in magnitude) to {\bf m12i}. However, as a lower-mass halo {\bf m11f} has a virial temperature too low to collisionally ionize \NeVIII, so in both MHD+ and CR+ runs it is photo-ionization dominated, and so the CR+ run produces an enhancement (rather than a decrease) in \NeVIII columns at larger radii $r \gtrsim 50\,$kpc akin to the enhancement in \OVI (owing to the cool gas being more diffuse and hence more easily-ionized).\footnote{While the choice of FG09 or HM12 UV backgrounds make relatively little difference for the \NeVIII in {\bf m12i} and {\bf m12m}, as it is collisionally-ionized, in {\bf m09}, {\bf m10q}, and {\bf m11f} the \NeVIII is photo-ionized in the MHD+ and CR+ runs, so the HM12 background predicts a factor $\sim 2$ higher \NeVIII column in each case.}

The dependence on redshift is also investigated by contrasting CR+ with MHD+ runs in {\bf m12z2}, a halo that reaches MW-mass $M_\mathrm{halo} = 1.7\times10^{12}\ M_\odot$ already at $z \sim 2$. Despite the fact that these are exactly the masses where we saw the most dramatic effects in {\bf m12i} and {\bf m12m}, we see very little effect of CRs on the columns at $z\sim2$. This is consistent with the analysis in \citet{hopkins2019but}, which showed that galaxy properties, star formation histories, and all other properties analyzed therein for these simulations were influenced by CRs only at $z\lesssim 1-2$; they also explicitly showed that MW-mass halos at $z\gtrsim 2$ have CR pressure insufficient to balance gravity. As discussed below, from simple analytic arguments we expect this, as a consequence of the much higher CGM densities, accretion, and cooling rates at high-$z$.

\section{Interpretation and Implications}
\label{sect:itp}

\subsection{Equilibrium Scalings}
\label{sec:discussion.scalings}

We showed above that effects of CRs are maximized at MW masses ($M_{\rm halo} \gg 10^{11}\,M_{\sun}$) at low redshifts ($z\lesssim 1-2$), where halo gas is supported by CR pressure instead of thermal pressure. A simple physical explanation for this is given in \citet{hopkins2019but}: if we assume CRs originate from a point-source galaxy (small compared to the CGM), transport with isotropically-averaged bulk transport speed $\tilde{v}_{\rm cr} \sim {\rm MAX}(\tilde{\kappa}/r,\,\tilde{v}_{\rm st})$ (where $\tilde{v}_{\rm st} \sim v_{\rm st}/3 \sim v_{A}/3$ and $\tilde{\kappa}\sim \kappa_{\|}/3$ represent the CR streaming speed and diffusion, respectively, as we model them), have negligible losses, and are in steady-state with an injection rate $\dot{E}_\mathrm{cr} = \epsilon_\mathrm{cr}\epsilon_\mathrm{SNe}\dot{M}_{\ast}$ (where in our simulations $\epsilon_\mathrm{cr} = 0.1$ and $\epsilon_\mathrm{SNe}\sim (10^{51}\ \mathrm{erg}/70 M_\odot)$), then the CR pressure is $P_{\rm cr}(r) = e_{\rm cr}(r)/3 \approx \dot{E}_{\rm cr}/12\pi\,\tilde{v}_{\rm cr}\,r^{2}$. This analytic estimation produces resonable CR pressure gradient profiles $\nabla P_\mathrm{cr}$ which are consistent with our simulation results, as shown in Fig. \ref{fig:grad_P}. For a CGM in an isothermal sphere obeying the usual virial scalings for dark matter, \citet{hopkins2019but} further showed that the corresponding ratio of the radial CR pressure gradient to gravitational force in a star-forming galaxy is:
\begin{align}
 \frac{|\nabla P_\mathrm{cr}|}{|\rho \nabla \Phi|} \sim \frac{0.5 \alpha\, \epsilon_\mathrm{cr}}{f_\mathrm{gas,0.1} \tilde{\kappa}_{29} (1+z)^{3/2}}\left(\frac{M_*}{f_b M_\mathrm{halo}}\right),
 \label{eq:depend_M_z}
\end{align}
where $\tilde{\kappa}_{29}\equiv \tilde{\kappa}/10^{29} \, \mathrm{cm^2\ s^{-1}}$, $f_\mathrm{gas,0.1}\equiv f_\mathrm{gas}/0.1$ is the mass ratio of gas to total matter in the halo, $f_b\equiv \Omega_b/\Omega_m$ is the universal baryon fraction, and $\alpha \equiv \dot{M}_{\ast} / (M_{\ast}/t_{\rm Hubble}[z]) \sim 1$ is a dimensionless specific star formation rate. \citet{hopkins2019but} shows this provides a reasonable approximation to $P_{\rm cr}(r)$ as a function of $M_{\rm halo}$ and $z$.

This shows that the importance of CRs in the CGM scales with the stellar-to-halo mass ratio $M_{\ast}/M_{\rm halo}$, which drops precipitously in dwarf galaxies: CRs are injected at too-low a rate (given lower $M_{\ast}$ and hence $\dot{M}_{\ast}$ and $\dot{E}_{\rm cr}$), they escape too effectively, and (in addition) there is no real hot, quasi hydro-static gaseous halo present for them to ``work upon.'' We also see the term $(1+z)^{3/2}$, which originates from the fact that the CGM is more dense at high redshifts ($\rho \propto (1+z)^{3}$, determined by the fast increase of the characteristic density of the CGM with redshift that is only partially compensated by a slower increase of the input rates of the CRs at a given halo mass with redshift), and shows that CRs should decrease in importance at high redshifts. 

Now consider a massive, low-$z$ halo where CR pressure balances gravity as shown in Fig.~\ref{fig:grad_P}, so $\rho\,\nabla\Phi \approx \rho\,V_{c}^{2}/r \approx \nabla P_{\rm cr}$, giving $\rho(r) \sim \rho_{\rm eq} \sim \dot{E}_{\rm cr} / 12\,\pi\,V_{c}^{2}\,\tilde{v}_{\rm st}\,r^{2}$. Gas with $\rho > \rho_{\rm eq}$ will sink, while that with $\rho < \rho_{\rm eq}$ will float, until it reaches the radius where $\rho \sim \rho_{\rm eq}$, {\it independent} of gas temperature. This estimate of $\rho_\mathrm{eq}$ matches the density profiles of CR+ simulations quite well, as shown in Fig. \ref{fig:profile_rhoT}. In contrast, in MHD+ runs where thermal pressure dominates, virial balance specifically implies a characteristic temperature $T(r) \sim T_{\rm eq} \sim \mu\,m_p\,V_{c}^{2} / k_{B}$, with small-scale fluctuations in local thermal pressure equilibrium ($\rho \propto T^{-1}$).

\subsection{Multiphase Structure}
\label{sec:discussion.multiphase}

\begin{figure*}
\begin{centering}
\includegraphics[width={\textwidth}]{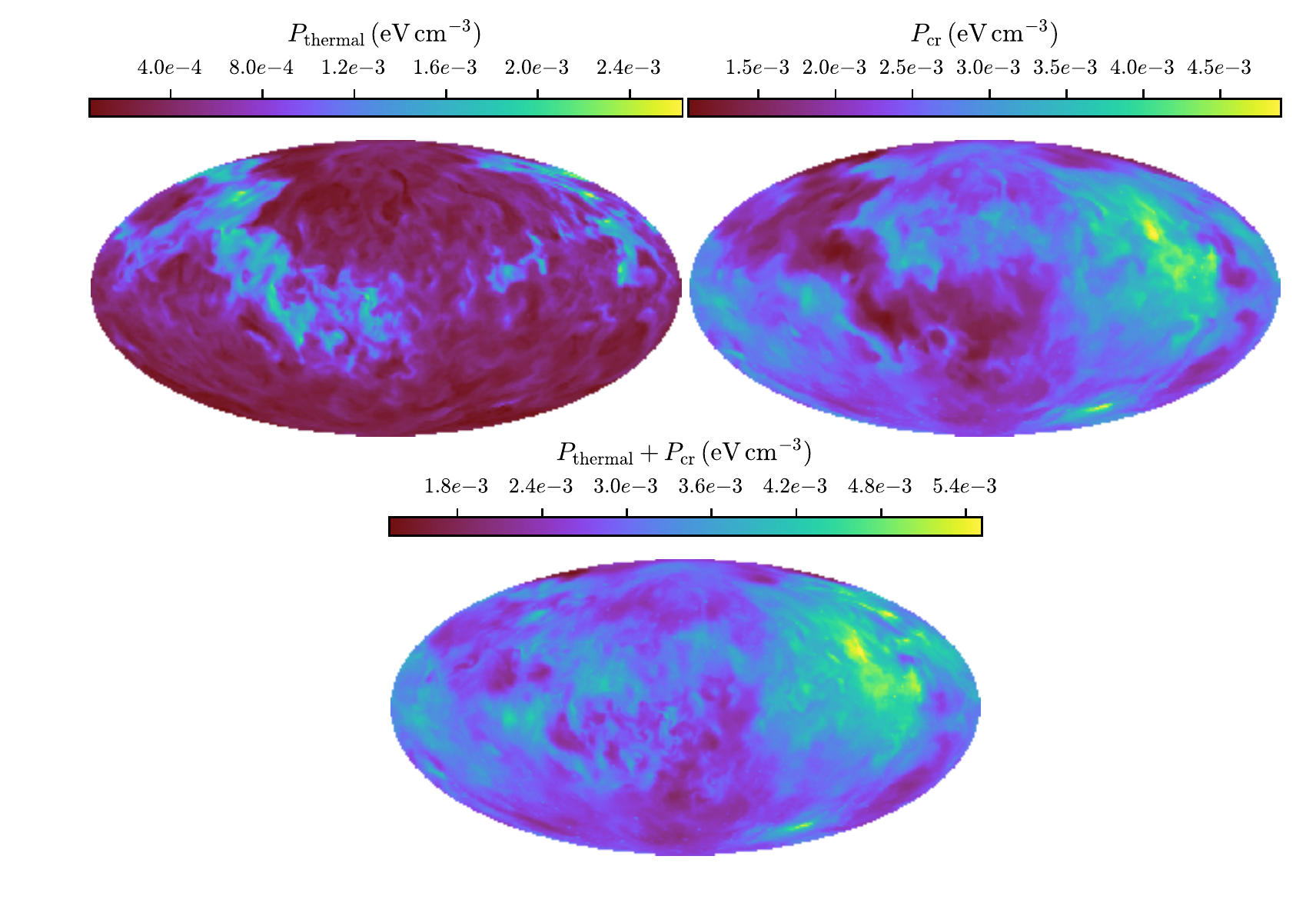}
\end{centering}
\vspace{-0.5cm}
\caption{Mollweide projections of gas thermal pressure {$P_\mathrm{thermal}$} and CR pressure $P_\mathrm{cr}$ in a narrow spherical shell at radius $r \approx 150\, \mathrm{kpc}$, for the {\bf m12i} CR+ run at $z=0$. {$P_\mathrm{thermal}$}  and $P_\mathrm{cr}$ are locally anti-correlated at a given $r$, such that the total pressure $P_{\rm tot}(r)$ is closer to uniform.
\label{fig:pressure_at_surface}\vspace{-0.5cm}}
\end{figure*}

In CR-dominated halos, note that the equilibrium density $\rho_{\rm eq}(r)$ defined above (\S~\ref{sec:discussion.scalings}) does not depend on gas temperature, which implies gas at $\rho \sim \rho_{\rm eq}$ at a given $r$ may well be at different temperatures. This is seen in Fig.~\ref{fig:profile.density}, where gas at orders-of-magnitude different temperatures co-exists at similar densities at the same radius $r$. In CR+ runs, the cool phase can therefore remain out of thermal pressure equilibrium with the hot medium: it prefers at $\sim \rho_{\rm eq}$, as compared to MHD+ cases where thermal pressure dominates so cool phase gas must be in local thermal pressure equilibrium (so $\rho_{\rm cool} \sim \rho_{\rm hot}\,(T_{\rm hot}/T_{\rm cool})$). Thus the CR+ runs can support (via CR pressure) a smoothly-distributed and volume-filling cool phase, consistent with the gas morphology plot in Fig.~\ref{fig:morph}, and the column density profiles in Fig.~\ref{fig:column_m12i}. Some qualitatively similar effects have been seen in idealized simulations of galactic outflows, as well \citep[see e.g.][]{girichidis2018cooler}. 

In greater detail, Fig.~\ref{fig:pressure_at_surface} considers local pressure fluctuations instead of the global pressure gradient, specifically comparing gas thermal, CR, and total pressure of gas in a narrow shell at a specific $r\approx 150\,$kpc, in the CR+ run. The pressure fluctuations are largely confined to the ``shell surface'' (i.e.\ directions tangential to $r$, not along $r$), so they do not alter the global radial force balance between CRs and gravity. The thermal pressure is largely below $10^{-3} \, \mathrm{eV\ cm^{-3}}$, corresponding to cool phase gas at a few $10^4\ \mathrm{K}$, though there are some regions where thermal pressure exceeds $10^{-3} \, \mathrm{eV \ cm^{-3}}$ in the warm phase with $T \gtrsim 10^{5}$\,K. The CR pressure is generally larger than thermal by a factor $\sim 10$, but in warm regions where thermal pressure peaks, CR pressures drop to a minimum value $\sim 1.5\times 10^{-3} \, \mathrm{eV\ cm^{-3}}$ comparable to or even less than the local thermal pressure. Fig.~\ref{fig:pressure_at_surface} implies that although the cool and hot phases are out of thermal pressure equilibrium, they are indeed in total pressure equilibrium with CR pressure taken into account, i.e.\ $P_{\rm thermal}(r) + P_{\rm cr}(r) \approx $\,constant at a given $r$. Quantitatively speaking, in Fig. \ref{fig:pressure_at_surface}, the standard deviation of the thermal pressure $\langle\delta P / P \rangle$ is $0.62$, while that of the total pressure is only $0.16$. As these fluctuations evolve, hot phase-gas carrying high thermal pressure rapidly flows into and becomes embedded in the ``troughs'' of CR pressure fluctuations, while cool phases (at $\rho_{\rm eq}$) are weakly ``squeezed'' by this hot gas while being smoothed by bulk CR pressure.

It is also useful to present some numbers regarding the total mass of cool gas, for the {\bf m12i} CR+ run as a representative MW-mass halo. The total cool ($10^4\ \mathrm{K} < T< 10^5\ \mathrm{K}$) gas masses in {\bf m12i} at $z\sim0.15$ are respectively $2.5\times 10^{10}\ M_\odot$ ($r < 100\ \mathrm{kpc}$), $3.7\times 10^{10}\ M_\odot$ ($r < 150\ \mathrm{kpc}$), $4.6\times 10^{10}\ M_\odot$ ($r < 200\ \mathrm{kpc}$) and $5.6\times 10^{10}\ M_\odot$ ($r < 300\ \mathrm{kpc}$). A significant amount of the halo baryon budget in this model resides at large scale, consistent with the observational claim that the inner halo of $\sim L_{\ast}$ galaxies is baryon deficient \citep{bregman2018extended}, and similar to observational estimates of a total gas mass $\gtrsim 10^{10}M_\odot$ within the virial radius (e.g., \citealt{stocke2013characterizing,werk2014cos,stern2016universal,prochaska2017cos,keeney2017characterizing}). The {\bf m12m} run,  which has a similar halo mass with {\bf m12i}, produces relatively less cool gas than {\bf m12i}, with a difference of $\sim8\%$ for $r < 100\ \mathrm{kpc}$ up to $\sim 30\%$ for $r < 300\ \mathrm{kpc}$. This indicates that although the qualitative properties of the CGM are similar in halos of the same mass at $z\sim0$, different accretion histories can introduce some variations. A detailed quantification of how the accretion history affects CGM properties is beyond the scope of this work.

\subsection{Photoionized \OVI \&\ Characteristic Columns}

We can write the \OVI number density as $n_\mathrm{OVI} = f_\mathrm{OVI}(n,T)\, (f_\mathrm{O}\, Z / \mu_\mathrm{O})  \, n_{\rm H}$, where $f_\mathrm{OVI}$ is the ionization correction (fraction of O in \OVI); $f_\mathrm{O}$, $Z$, and $\mu_\mathrm{O}$ are the oxygen mass fraction per unit metal mass, metallicity, and dimensionless mean molecular weight of O (so $f_\mathrm{O}\, Z / \mu_\mathrm{O}$ is the number of O nuclei per nucleon), and $n_{\rm H} = \rho/m_{p}$ is the nucleon number density. For the MW-mass, CR+ runs, the median gas density and temperature run from $\sim (10^{-4}\,{\rm cm^{-3}}\, , \, 3\times10^{4}\,{\rm K})$ at $r\sim 50\,$kpc to $\sim (10^{-5}\,{\rm cm^{-3}}\, , \, 10^{5}\,{\rm K})$ at $r\sim 200\,$kpc roughly following $n\propto r^{-1.5}$ (Fig.~\ref{fig:profile_rhoT}) and it is this gas which contributes most of the \OVI (Fig.~\ref{fig:phase_ion}). Along this temperature-density range, for the UV background at $z\lesssim 0.5$, the value of $f_\mathrm{OVI}$ given by photo-ionization is approximately constant at $f_\mathrm{OVI}\sim0.1$ (see e.g.\ Fig.~3 of \citealt{stern2018does}). If we assume solar abundance ratios ($f_\mathrm{O} / \mu_\mathrm{O} \sim 5\times10^{-4}$) and $Z\sim 0.1\,Z_{\odot}$ (representative of the outer halo with $r\gtrsim 100\,$kpc), and integrate\footnote{We integrate between radii where $\sim 10^{-5}\,{\rm cm^{-3}} < n < 10^{-4}\,{\rm cm^{-3}}$, i.e.\ $50\,{\rm kpc} \lesssim r \lesssim 200\,{\rm kpc}$, as outside this range the photo-ionized fraction $f_\mathrm{OVI}$ drops rapidly.} to obtain the column density $N_\mathrm{OVI} = \int n_\mathrm{OVI}\,d\ell \sim 2\,n_\mathrm{OVI}\,R$ along a median sightline at impact parameter $R$, we obtain a characteristic column $N_\mathrm{OVI}  \sim 10^{14.5}\,{\rm cm^{-2}}\,(R/100\,{\rm kpc})^{-0.5}$. Both the typical value, and fairly flat profile (falling by just a factor $\sim 2$ from $\sim 50-200\,$kpc) is broadly consistent with our \OVI column density profile in Fig.~\ref{fig:column_m12i}.

Scaling this to other halos, assuming the gas lives along the $\rho_{\rm eq}$ value given by equilibrium with CR pressure, predicts a relatively weak dependence of the $N_\mathrm{OVI}$ on halo mass, until (at dwarf masses) the density $\rho_{\rm eq}$ falls below the minimum density $\sim 10^{-5}\,{\rm cm^{-3}}$ at all radii, at which point $f_\mathrm{OVI}$ and $N_\mathrm{OVI}$ fall rapidly. This occurs for our {\bf m11b} and {\bf m10q} runs (compare Figs.~\ref{fig:profile_rhoT}, \ref{fig:column_halos}). 

The characteristic density $\lesssim 10^{-4}\,{\rm cm^{-3}}$ and distances/path-lengths $\sim 100\,$kpc of the \OVI gas therefore naturally result in the CR+ runs from primarily photo-ionized gas (as argued by \citealt{stern2018does} based on the observed \HI associated with the observed \OVI), supported by CR pressure. It is worth noting that ``maintaining'' the \OVI observed is energetically demanding if it is collisionally ionized \citep[see][]{lehner2014galactic,oppenheimer2017flickering,mcquinn2018implications,stern2018does,stern2019cooling,ji2019simulations}, as it must be kept near the peak of the cooling curve. 

The CR-dominated scenario may also explain the observed differences in \OVI between star-forming and non-star-forming galaxies: in steady state, the CR pressure is proportional to the CR injection, hence star formation rate, so the CR pressure becomes subdominant around galaxies in which star formation is quenched. Therefore, in order to maintain a high thermal pressure, the halo gas must be either hot and diffuse, or cold and dense. In either case, the gas may not be able to produce sufficient \OVI column densities via either photoionization or collisional ionization. However, we caution that if quenching owes to AGN producing CRs \citep{su2019failure}, the above considerations may not apply. 

Some previous studies have argued against a photo-ionization origin for \OVI, but in fact our CR+ simulations here naturally explain the supposedly problematic observations. For example, \citet{werk2016cos} argue that reproducing the \NV/\OVI ratios requires path-lengths $\sim 100\,$kpc, which they assumed were unphysically large. But our CR+ simulations do reproduce the observed \NV/\OVI ratios (Fig.~\ref{fig:column_m12i}), and indeed {\it do} have path lengths reaching $\gtrsim 100-200\,$kpc, so these are clearly not unphysical. There are several factors of $\lesssim 2$ (owing to e.g.\ slightly different metallicity, temperature structure, and UV background shape, compared to that assumed in \citealt{werk2016cos}) which reduce the required path lengths for \OVI in our simulations. But the most important distinction is that \citet{werk2016cos} implicitly assumed \OVI would be in local thermal pressure equilibrium, requiring it be confined to much smaller, denser structures (as in our MHD+ runs), while in our CR+ runs the fact that it is {\it not} in thermal pressure equilibrium means it can be diffuse and volume-filling over a $\gtrsim 200\,$kpc radius (Fig.~\ref{fig:profile.density}).  Moreover, the diffuse, volume-filling nature of the \OVI gas on scales $\gtrsim 150\,$kpc naturally explains the ``broad'' linewidths of $\gtrsim 40-100\,{\rm km\,s^{-1}}$ seen in the high-column \OVI absorbers: these are tracing the bulk motions of the gas (inflow/outflow/mergers/sloshing/rotation/etc.), just like the usual assumption for models of volume-filling collisionally ionized gas -- the \OVI\ is not coming from dense clouds or filaments. A more detailed study of the absorber kinematics will be the subject of future work.

\section{Discussion \&\ Conclusions}
\label{sect:diss}

We have investigated the impact of cosmic rays on the CGM in FIRE-2 simulations, comparing fully-cosmological magnetohydrodynamic (``MHD+'') simulations including anisotropic conduction and viscosity and detailed models for for the ISM, cooling, star formation, and stellar feedback, to models including these physics as well as explicit CR transport and gas coupling (including anisotropic streaming and diffusion, collisional and streaming losses, and injection in SNe). In this study, we analyze six representative halos from the FIRE-2 suite, with halo masses ranging from $\sim 10^{10}M_\odot$ to $10^{12}M_\odot$. We have in previous work ensured that the simulations with CRs (``CR+'') reproduce observational constraints on CR populations in both the MW and nearby galaxies. Here, we explore the physical state of the CGM, with and without CRs, in halos from ultra-faint dwarf through MW masses. We construct mock observations of absorption-line profiles to compare directly with observed column density distributions of various ions as a function of impact parameter from galaxies. 

\subsection{Conclusions}

Our main conclusions are as follows. 

\begin{enumerate}

  \item CR pressure can dominate the CGM pressure at radii $\sim 30-300\,$kpc, supporting the gas. For MW-mass halos, the CR pressure gradient in our CR+ runs is larger than the thermal pressure gradient by more than one order of magnitude in the CGM, and dominates the support of the gas against gravity. The effect of CRs shows a strong halo mass dependence: CRs are negligible in the CGM of dwarfs ($M_{\rm halo} \lesssim 10^{11} \ M_\odot$), but become important in intermediate-mass halos (a few $10^{11} \ M_\odot$), and dominant in the MW-mass halos ($\sim 10^{12} \ M_\odot$).

  \item In CR pressure-dominated halos, CRs can change the phase and morphology of the halo gas dramatically. In MW-mass halos, the dominant gas temperature decreases from $\gtrsim 10^5 \, \mathrm{K}$ (in our MHD+ runs) to a few $10^4 \, \mathrm{K}$ (in our CR+ runs). In the CR+ runs, the hot ($T \gtrsim 10^6\ \mathrm{K}$) gas in the halo nearly disappears, and the cool ($T\sim$ a few $10^4\ \mathrm{K}$) gas becomes volume-filling (as compared to tightly-confined in dense clumps or filaments, in the MHD+ runs).

  \item An equilibrium gas density $\rho_{\rm eq}$ is implied by balancing CR pressure support and gravity in CR pressure-dominated halos. This equilibrium density is independent of the gas temperature, and indeed we show that halo gas with orders-of-magnitude different temperatures can co-exist at similar densities (at a given galacto-centric radius) in our CR+ runs. As a consequence (because the CRs provide most of the pressure), the cool phase appears ``under-pressured'' -- i.e.\ out of local {\it thermal} pressure equilibrium with the warm/hot phases ($P_{\rm thermal,\,cool} \ll P_{\rm thermal,\,hot}$, although we show it is in approximate {\it total} pressure equilibrium). This also means that the density structure of the CGM is ``smoother'' in CR+ runs, since cool gas can be volume-filling as opposed to confined (as noted above).

  \item In CR+ runs, the CR pressure and gas pressure are locally anti-correlated, with the total pressure roughly constant in spherical shells in the CGM at a given galacto-centric radius. Thus halo gas is in both local total pressure balance as well as global (virial) pressure balance in the halo, but only when CR pressure is included. CR pressure is therefore dominant in cool-phase gas (making up for its low thermal pressure), but becomes increasingly sub-dominant in warm/hot phases.

  \item In CR pressure-dominated, MW-mass halos, all ions (\HI--\NeVIII) observed by COS are primarily photo-ionized, including \OVI and \NeVIII which are predominantly collisionally-ionized in the MHD runs. As a consequence of this and the overall shift in the phase/temperature of the gas, CRs effectively enhance the column densities of the low and mid ions contributed by the photo-ionized cool gas, and reduce the column densities of the high ions owing to the decrease of collisionally-ionized hot gas. The CR+ runs yield \OVI, \HI and \NV columns qualitatively consistent with observational detections and limits, in contrast with the MHD runs which under-predict \OVI and \HI. The CR+ runs however over-predict the observed \SiIV columns at $\sim 50\,\mathrm{kpc}$.  In addition, in CR+ runs, the sightline-to-sightline scatter in the low/mid ion column densities becomes much smaller, owing to the change in phase structure (with cool gas more volume-filling and at closer to uniform densities $\sim \rho_{\rm eq}$, as opposed to concentrated in much denser clumps with low volume-filling factors).  
  
  \item For MW-mass halos, \OVI comes primarily from collisional ionization of warm gas around $3\times 10^5 \, \mathrm{K}$ in our MHD+ runs, but from photo-ionization of cool gas around and below $10^5 \, \mathrm{K}$ in our CR+ runs. When CR pressure is dominant, given the UV background model adopted here, typical \OVI column densities are $\gtrsim 10^{14.5} \, \mathrm{cm^{-2}}$ at impact parameters out to $\gtrsim 150 \, \mathrm{kpc}$, consistent with observations of low redshift SF galaxies, as the gas number density remains near and above $10^{-5} \, \mathrm{cm^{-3}}$ at which the photo-ionized \OVI fraction peaks. We caution that the absolute value of the \OVI and \NeVIII columns is sensitive to the shape of the UV background assumed, when photo-ionization dominates the observed ions, but our qualitative conclusions about the effects of CRs remain robust to the exact choice of UV background.

\end{enumerate}

\subsection{Caveats}

This work is subject to a number of caveats due to both observational and theoretical uncertainties.

\begin{enumerate} 

\item{{\it CR transport physics}: The major theoretical uncertainty here comes from assuming a spatially and temporally-constant effective CR diffusion coefficient  $\kappa_{\|}$ (with streaming at $\sim v_{A}$) in our simulations. \citet{hopkins2019but} showed that much lower $\kappa_{\|}$, for example, can produce much weaker effects of CRs in galaxies. While the value of $\kappa_{\|}$ adopted here is calibrated to reproduce CR population modeling in the MW and $\gamma$-ray observations in nearby galaxies \citep{chan2019cosmic,hopkins2019but}, these observations only really constrain the value of $\kappa_{\|}$ in the ISM and ``inner CGM'' ($\lesssim 10\,$kpc) at $z\sim 0$. It is plausible that the CR diffusivity increases rapidly as CRs propagate into the CGM (with lower gas densities and magnetic fields), allowing CRs to freely escape instead of being confined inside halos as they are in our CR+ runs. The median diffusivity may also evolve in time: at high redshifts merger and inflow/outflow activity could enhance turbulence and modify CR diffusion \citep{beresnyak2011numerical}. Better observational and theoretical constrains on CR propagation are under active investigation and will definitely benefit future work.}

\item{{\it Resolution and other small-scale physics}: Our simulations might not fully capture some interesting small-scale processes. For example, CRs may modify un-resolved hydrodynamic instabilities \citep{suzuki2014linear} and thermal instabilities, and the propagation of CRs can be altered by the micro-scale structure and damping of MHD turbulence \citep{lazarian2016damping,xu2016damping}. Even in hydrodynamic cases, various authors have argued that higher resolution allows better resolution of thermal instabilities and can therefore be especially important for resolution of cool ions \citep{mccourt2017characteristic,peeples2018figuring,hummels2018impact,van2018cosmological}. A detailed study of the effects of resolution on our CGM predictions will be presented in future work (Hummels et al., in prep.); however we have compared our MW-mass halos with and without CRs at mass resolution levels $m_{i}/M_{\odot} \sim (7000,\, 56000,\ 450000)$ and find that while there is some resolution dependence to our predictions, the {\it systematic} effect of CRs, and essentially all of our predictions here, are similar at all resolution levels. And for our dwarfs we have compared simulations at resolution $m_{i}/M_{\odot} \sim (250,\, 2100,\ 16000)$, over which range we see negligible resolution dependence in either MHD+ or CR+ runs. Both of these conclusions are also consistent with the comparison of resolution effects on galaxy properties in the MHD+ and CR+ runs presented in \citet{chan2019cosmic} and \citet{hopkins2019but}.}

\item{{\it Hot components and other CGM observables}: In our CR-dominated MW-mass halos, the hot phase with $T\gtrsim 10^6 \, \mathrm{K}$ (and \NeVIII) is strongly suppressed at radii $50\ \mathrm{kpc}< r < 200 \, \mathrm{kpc}$, while some X-ray observations suggest the existence of such a hot phase at $\sim 10^6 \, \mathrm{K}$ in the MW halo (e.g, \citealt{fang2012hot}). But it is also possible that the MW X-rays might come from small scales $\lesssim 20\,\mathrm{kpc}$ (see \citealt{bregman2018extended,stern2019cooling}), or be generated by diffuse gas with very little mass (hence, un-resolved in our fixed mass-resolution simulations) in hybrid CR/thermal winds \citep{everett2008milky}. A quantitative comparison to X-ray observations and other CGM observables (including e.g.\ the SZ effect, absorber kinematics, emission, and more) is clearly needed.}

\item{{\it AGN}: We have neglected AGN in our study here, since the main focus of this paper is the galaxy halos with halo mass $\lesssim 10^{12}\,M_{\odot}$. In massive ($\gg 10^{12}\,M_{\odot}$) halos, AGN jets or ``bubbles'' can contain large CR energies (orders-of-magnitude larger than those produced by SNe) and AGN feedback may well dominate the dynamics of the CGM: we are exploring this in parallel work \citep[see e.g.][]{su:turb.crs.quench,su2019failure}. One important conclusion from this work and \citet{hopkins2019but} is that CRs from SNe alone become {\em less} significant in halos with masses much larger than MW-like, owing to the larger CGM pressure and higher temperatures, and their lower SFR (hence SNe) rates. For simple energetic reasons (given the observed BH-host galaxy scaling relations; see \citealt{kormendy:2013.mbh.mgal.review}), it is generally believed that in the lower-mass halos studied here, AGN feedback is sub-dominant, but this remains to be tested.}

\end{enumerate}

Therefore, we can {\it not} assert that the CRs {\it must} necessarily exert a major influence on the CGM -- it is perfectly plausible that reality lies ``in between'' our MHD+ and CR+ runs. Instead, we argue that CRs {\it could} be essential to the physics of the CGM, and propose a model of the CGM which is dominated by CR pressure which makes a number of distinct predictions.

\acknowledgments{We thank the anonymous referee for a constructive and insightful report which improved our paper. SJ is supported by a Sherman Fairchild Fellowship from Caltech.  SJ thanks Joe Burchett, Zheng Cai, Taotao Fang, Peng Oh, J.\ Xavier Prochaska, Gwen Rudie, Mateusz Ruszkowski, Britton Smith, Daniel Wang and Jessica Werk for many helpful comments and discussions, and the Aspen Center for Physics supported by NSF PHY-1607611 for its hospitality where part of this work was completed. Support for PFH and co-authors was provided by an Alfred P. Sloan Research Fellowship, NSF Collaborative Research Grant \#1715847 and CAREER grant \#1455342, and NASA grants NNX15AT06G, JPL 1589742, 17-ATP17-0214. DK was supported by NSF grant AST-1715101 and the Cottrell Scholar Award from the Research Corporation for Science Advancement. CAFG was supported by NSF through grants AST-1517491, AST-1715216, and CAREER award AST-1652522, by NASA through grant 17-ATP17-0067, by STScI through grants HST-GO-14681.011, HST-GO-14268.022-A, and HST-AR-14293.001-A, and by a Cottrell Scholar Award from the Research Corporation for Science Advancement. Numerical calculations were run on the Caltech compute cluster ``Wheeler'', allocations from XSEDE TG-AST120025, TG-AST130039 and PRAC NSF.1713353 supported by the NSF, and NASA HEC SMD-16-7592. We have made use of NASA's Astrophysics Data System. Data analysis and visualization are made with {\small Python 3}, and its packages including {\small NumPy} \citep{van2011numpy}, {\small SciPy} \citep{oliphant2007python}, {\small Matplotlib} \citep{hunter2007matplotlib}, {\small Healpy} \citep{2005ApJ...622..759G,Zonca2019}, the {\small yt} astrophysics analysis software suite \citep{turk2010yt} and the absorption spectra tool {\small Trident} \citep{hummels2017trident}, as well as the spectral simulation code {\small CLOUDY} \citep{ferland20172017}.}\\

\dataavailability{The data supporting the plots within this article are available on reasonable request to the corresponding author. A public version of the GIZMO code is available at \gizmourl.

Additional data including simulation snapshots, initial conditions, and derived data products are available at \FIREurl.}\\

\vspace{-0.2cm}
\bibliography{ms_extracted}

\appendix

\section{CGM Pressure Support in Additional MW-mass Systems}
\label{apdx:pressure_m12}

\begin{figure*}
    \begin{centering}
    \includegraphics[width={\textwidth}]{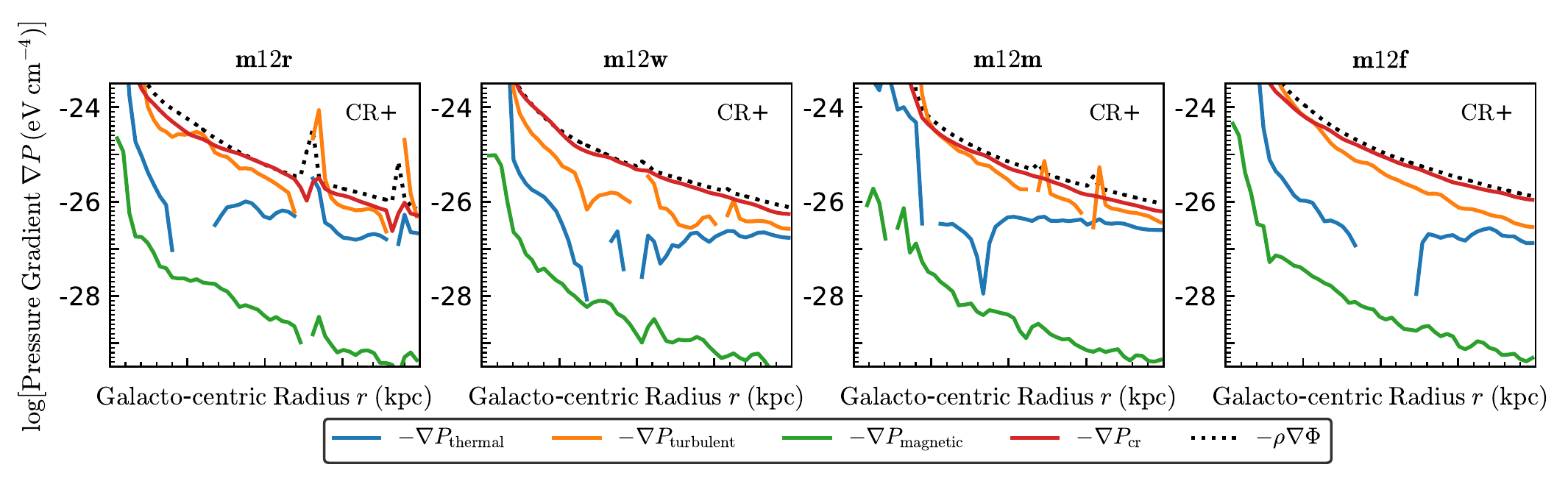}
    \end{centering}
    \vspace{-0.6cm}
    \caption{Radial pressure profiles, as Fig. \ref{fig:grad_P}, for CR+ runs of additional MW-halos {\bf m12r}, {\bf m12w}, {\bf m12m} and {\bf m12f} with halo masses of $8.9\times10^{11}M_\odot$, $1.0\times10^{12}M_\odot$, $1.5\times10^{12}M_\odot$ and $1.6\times10^{12}M_\odot$ respectively. As for {\bf m12} analyzed in the main text, the CR pressure gradient (red) dominates over thermal pressure gradient and roughly balances gravity in each of these MW-mass halos.
    \label{fig:grad_P_m12}}
    \end{figure*}

In the main text, we presented one halo ({\bf m12i}) of mass similar to the Milky Way and showed that its CGM is dominated by CR pressure. 
Here we present a larger sample of MW-mass halos to demonstrate that results obtained for {\bf m12i} regarding CR pressure are broadly generic at this mass scale. Figure \ref{fig:grad_P_m12} shows radial profiles of pressure gradients in additional MW-mass halos ({\bf m12r}, {\bf m12w}, {\bf m12m} and {\bf m12f}) from \citet{hopkins2019but}. In all of these MW-mass systems, CR pressure gradients (red) are overall the largest pressure gradients in the CGM, and are comparable with gravitational force. 
I.e., the gaseous halos are consistent with being predominantly supported by CR pressure. 
In all cases, the thermal pressure gradients are substantially subdominant.

\section{Systematic Effects of the UV Background Model: HM12 versus FG09}
\label{apdx:hm_vs_fg}

\begin{figure*}
\begin{centering}
\includegraphics[width={\textwidth}]{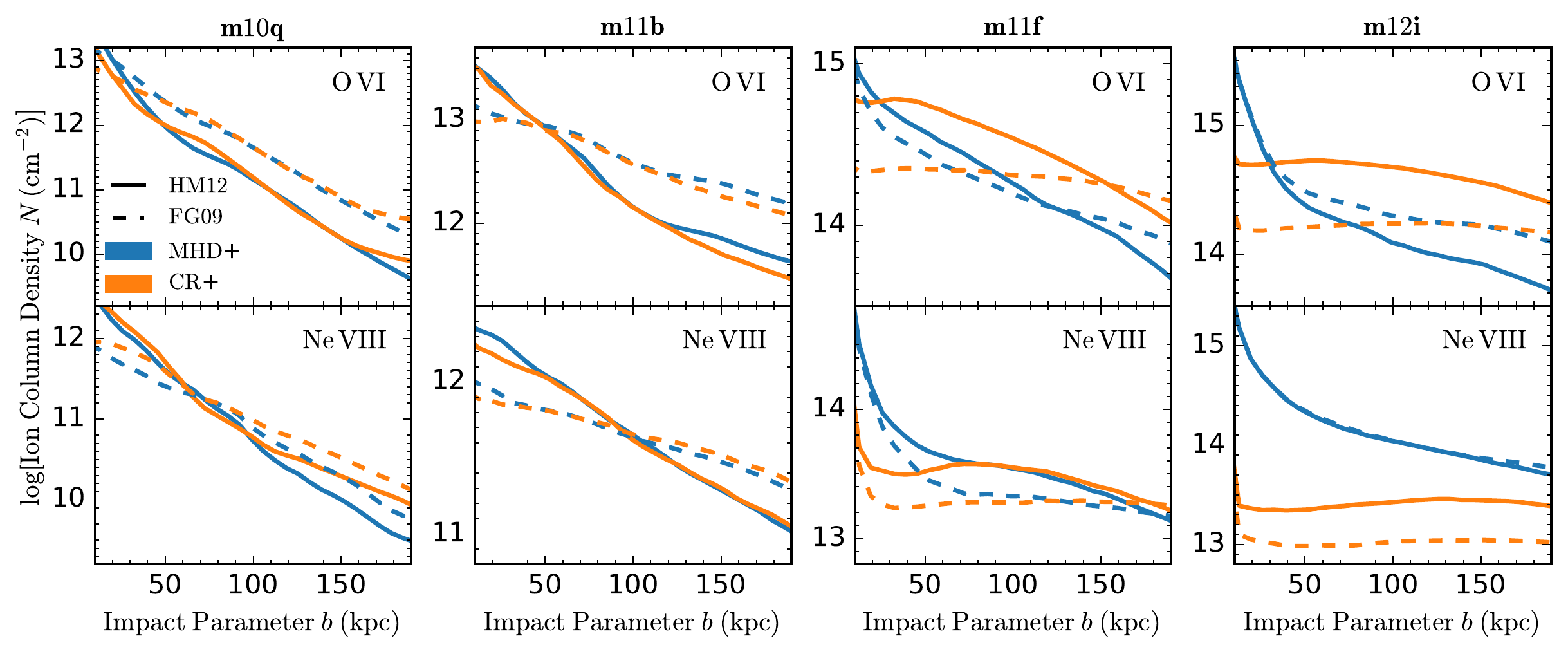}
\end{centering}
\vspace{-0.6cm}
\caption{Column density profiles of \OVI (top) and \NeVIII (bottom), assuming the UVB model from HM12 ({\it solid}) or FG09 ({\it dashed}), for MHD+ ({\it blue}) and CR+ ({\it orange}) runs at different halo masses (labeled). 
At all masses, the choice of UVB {\it systematically} shifts the predicted \OVI and \NeVIII columns. In general, FG09 produces flatter profiles with smaller/larger columns at smaller/larger $b$. However, our qualitative conclusions are robust. 
In dwarfs ({\bf m10q}, {\bf m11b}) there is no significant difference between CR+ vs.\ MHD+ runs, regardless of which UVB is adopted. 
At MW-mass, the qualitative differences between MHD+ and CR+ runs are robust to the UVB choice, but the effect of the UVB on mid/high ions is more dramatic in the CR+ runs because these are more photo-ionization-dominated (as opposed to collisionally-ionized). In particular, for CR+ runs the softer FG09 UVB produces less-efficient photo-ionization of \OVI and \NeVIII, reducing the predicted columns by factors $\sim 2-3$ at $b\lesssim 150\,$kpc compared to HM12.
\label{fig:column_halos_hm_fg}}
\end{figure*}

In this appendix, we compare two UVB models discussed in the paper, HM12 and FG19, for \OVI~and \NeVIII~columns. As discussed in \S \ref{sect:column}, these two models they differ in their treatment of the higher-energy EUV spectrum, which in turn leads to different predictions for the photo-ionization of mid/high ions. Fig.~\ref{fig:column_halos_hm_fg} presents \OVI and \NeVIII column densities produced assuming the HM12 or FG09 backgrounds, respectively, from both MHD+ and CR+ runs with various halo masses. In all cases, there is a systematic difference in the predictions for \OVI and \NeVIII depending on the UVB spectrum we assume: FG09 generally produces shallower profiles for mid/high ions, compared to HM12. 

In dwarfs there is no systematic difference between MHD+ and CR+ runs, for either HM12 or FG09 backgrounds: in other words our statement from the text that the effects of CRs in low-mass halos is negligible is robust. In more massive halos, the qualitative effects of CRs are similar independent of UVB, but we see the difference between the two UVB models in the CR+ runs is more pronounced than in the MHD+ runs (as we showed above, the CR+ runs are more strongly dominated by photo-ionization). For example, the \OVI column densities in the {\bf m12i} MHD+ run generally agree between HM12 and FG09 because they are collisionally-dominated, but in the CR+ run HM12 predicts a factor $\sim 3$ larger inner-halo \OVI columns compared to FG09. 

\begin{figure}
\begin{centering}
\includegraphics[width={0.5\textwidth}]{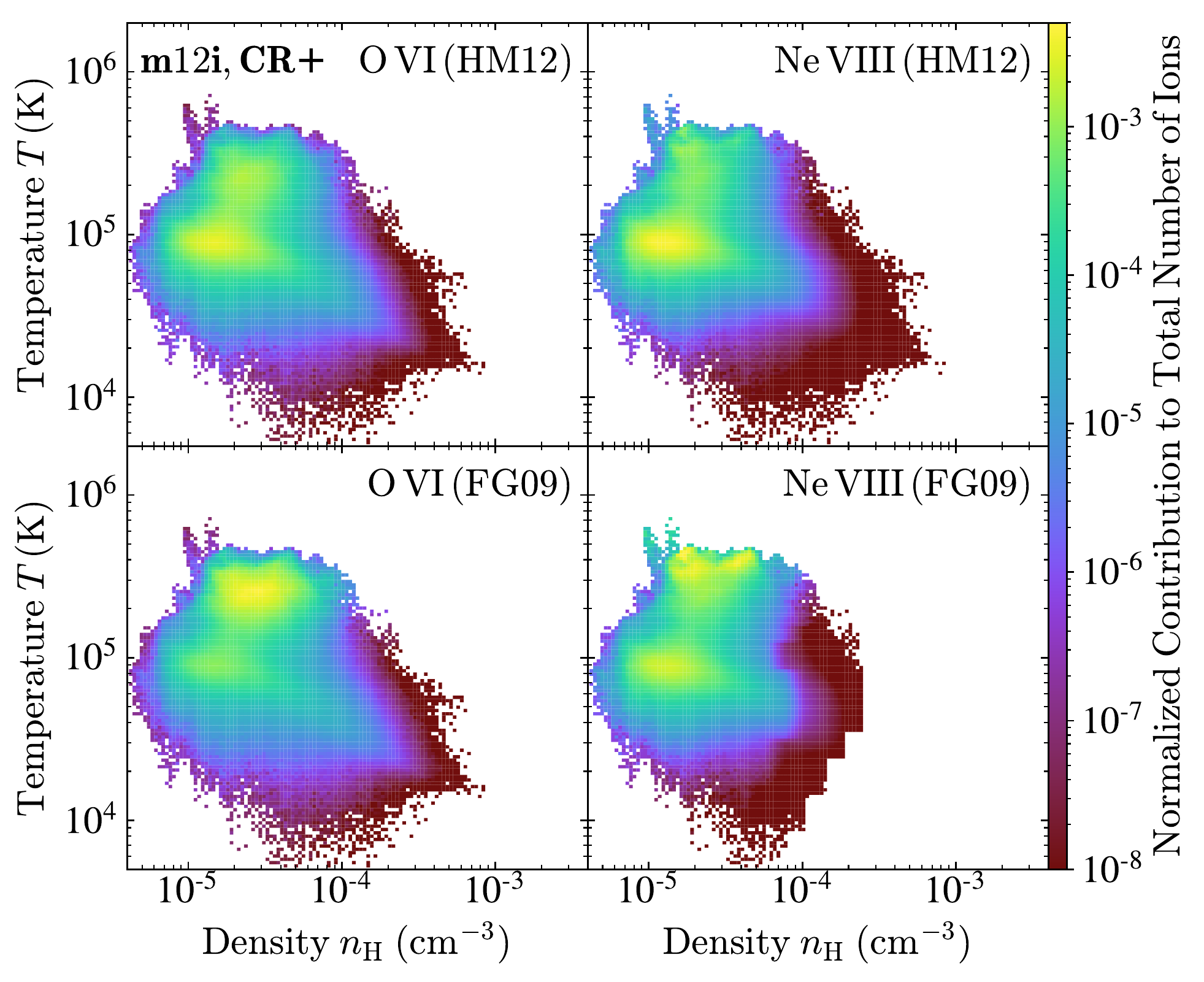}
\end{centering}
\vspace{-0.5cm}
\caption{Density-temperature diagrams of {\bf m12i}, CR+ for CGM ($50\,{\rm kpc}<r<200\,{\rm kpc}$) gas at $z=0$, weighted by the number of \OVI ({\it left}) and \NeVIII ({\it right}) ions. The ions are generated with both default HM12 (top) and FG09 (bottom). With FG09, the number of ions of both species coming from photo-ionized gas ($T\lesssim10^5\,\mathrm{K}$) decreases significantly. Thus the quantitative amount of  \OVI and \NeVIII produced from photo-ionization of the cool, diffuse CGM is sensitive to the UVB shape.
\label{fig:phase_ion_hm_fg}}
\end{figure}

In Fig.~\ref{fig:phase_ion_hm_fg}, we focus on the difference between HM12 and FG09 in ionizing \OVI and \NeVIII in the {\bf m12i} CR+ runs via density-temperature diagrams. The contribution to \OVI and \NeVIII coming from gas with $T\lesssim 10^5\,\mathrm{K}$ and $n_\mathrm{H}\sim$ a few $10^{-5}\,\mathrm{cm^{-3}}$ (where the majority of CGM gas in this run lies; see Fig.~\ref{fig:phase_dens_temp}) is reduced by a factor $\sim 2-3$ in the FG09 UVB model relative to HM12. This is, as we emphasized in the text, also similar to the density and temperature range where photo-ionization is most efficient.  Thus, if the mid/high CGM ions are primarily photo-ionized, care is needed in the treatment of the high-energy component of the UVB.
The more recent UVB model of \cite{faucher2020cosmic}, which includes a more detailed treatment of AGN than FG09, produces a high-energy spectrum in good agreement with HM12. We therefore favor the HM12 predictions for ions sensitive to photons of energy $>$4 Ry. 

\end{document}